# Detecting of photons in optical fields of complicated spatial structure

V.V. Klimov
P.N. Lebedev Physical Institute, Russian Academy of Sciences, 53
Leninsky Prospekt, Moscow 119991, Russia; e-mail : vklim@sci.lebedev.ru
D. Bloch, M. Ducloy
Laboratoire de Physique des Lasers, UMR 7538 du CNRS et de
l'Universite Paris13, 99 Avenue J-B. Clement F 93430
Villetaneuse, France. E-mail : bloch@lpl.univ-paris13.fr
J.R.Rios Leite
Departamento de Fisica Universidade Federal de Pernambuco,
50670-901 Recife, PE, Brazil
pacs 42.50.Tx 32.90.+a 33.80.-b

Abstract

We show that a photon detector, sensitive to the gradients of electromagnetic fields, or to magnetic fields, should be a useful tool to characterize the quantum properties of spatially-dependent optical fields. We discuss the excitation of an atom on a magnetic dipole or electric quadrupole transitions with the spiral photons of a Laguerre-Gauss (LG) or Bessel beams. We show that these beams are not true hollow beams, due to the presence of magnetic fields and gradients of electric fields on beam axis. This approach paves the way to an analysis at the quantum level of the spatial structure and angular momentum properties of singular light beams.

## 1. **Introduction**

In numerous approaches, the properties of a light field are connected to the properties of its electric field. This restriction still holds in most works of quantum optics, including all works following Glauber's theory about the quantum properties of optical fields and the detection of photons. Indeed, in his seminal paper Glauber (1963) justifies that in optics, one can restrict in most cases to a detector only sensitive to the electric field amplitude. The detector he considers, assumed to be of negligible size and extra wide frequency band, is implicitly an electric dipole (E1) detector. Although Glauber (1963) explicitly mentions that the possibility of correlation functions for the magnetic field could prove "someday useful", we are not aware of further works along this direction. Here, we establish that the consideration of the electric field only is restrictive when the spatial structure of optical fields becomes very complicated. In particular, we show that with suitable detectors, simply based upon atomic absorption on a magnetic dipole transition (M1) or on an electric quadrupole transition (E2), photons should be detectable in space regions where a standard E1 detector is unable to perform the detection.

Realizations of optical fields with complicated spatial structure are now numerous owing to the fast pace of development of nanotechnologies, as there are found with the optical fields near nanostructures, and metamaterials (Lee et al 2006). The spatial structure of electric fields between plasmonic nanoparticles is especially complicated (Klimov Guzatov 2007 ). However, connecting quantum properties of light with the specific structure of the field is simpler for a light field freely propagating in vacuum, than when the properties of a neighboring material medium have to be taken into account.

Another example of optical fields with complex spatial structure is the Laguerre-Gauss (LG) beams (Allen et al 1992, Allen, Padgett Babiker 1999, Allen Barnett, Padgett 2003, Barnet Allen 1994, Santamato 2004). In the present work, we analyze specifically these fields. Their behavior is typical of the one of spiral beams, that include the family of Bessel beams (Barnet Allen 2004). They are usually considered to be hollow beams, with a phase singularity, and no photon is expected to be detected on-axis with a standard detector relying on a E1 process. However, we will show that these beams are not truly hollow, as a detector of the magnetic field (M1 transition) or of the gradient of the electric field (E2 transition) yields an on-axis response. Moreover, LG beams have attracted a lot of interest owing to their orbital angular momentum (Allen 1992,1999; Barnet Allen 1994;Santamato 2004), and we show here that the transfer of more than one unit of angular momentum can be simply detected when the photon detection is operated on a higher than E1 transition.

The rest of the paper is as follows. In Section 2 we will present the theory of excitation of arbitrary quantum system by optical fields with arbitrary space structure. In section 3 we consider spatial properties of electric and magnetic fields and their gradients for LG beams. In section 4 we investigate interaction of such beams with elementary quantum systems which are characterized by electric dipole, magnetic dipole and electric quadrupole transition amplitudes. In section 5 we discuss results obtained.

## *2. General theory of Excitation of elementary quantum systems by optical fields of complicated spatial structure*

To calculate excitation rate of elementary quantum system by optical field with arbitrary space structure we will start with minimal coupling Hamiltonian with Coulomb gauge

$$\begin{aligned} H_{\text{int}} &= -\frac{e}{mc}\hat{\mathbf{A}}(\hat{\mathbf{r}})\hat{\mathbf{p}} + \frac{e^2}{2mc^2}\hat{\mathbf{A}}^2(\hat{\mathbf{r}}) \\ div\hat{\mathbf{A}}(\hat{\mathbf{r}}) &= 0 \end{aligned} \tag{1}$$

Here $\hat{\mathbf{p}}$ is operator of electron momentum and $\hat{\mathbf{A}}(\mathbf{r})$ is operator of vector potential at the electron position $\mathbf{r}$. The last term in (1) gives no contribution when calculating excitation rates.

The dipole approximation of Hamiltonian (1) is often used, that is why they substitute $\hat{\mathbf{A}}(\hat{\mathbf{r}}) \to \hat{\mathbf{A}}(\mathbf{r}_0)$ where $\mathbf{r}_0$ is radius vector of atom or molecule center mass. However in more complicated optical fields (that is for optical beams with singularities and for nanoscale optical fields) one should take into account further multipole expansion terms. Here we restrict ourselves to terms next to electric dipole terms, that is to magnetic dipole and electric quadrupole terms. In this case Hamiltonian will look like

$$H_{\text{int}} = -\frac{e}{mc}\hat{\mathbf{A}}(\mathbf{r}_0)\hat{\mathbf{p}}(\mathbf{r}) - \hat{\mathbf{H}}(\mathbf{r}_0)\hat{\mathbf{m}} - \frac{\partial \hat{A}_i(\mathbf{r}_0)}{\partial r_{0,j}}\hat{\tilde{Q}}_{ij} \tag{2}$$

where

$$\hat{\mathbf{H}}(\mathbf{r}_0) = rot\hat{\mathbf{A}}(\mathbf{r}_0) \tag{3}$$

is operator of magnetic fields and

$$\hat{\mathbf{m}} = \frac{e}{2mc}[\hat{\mathbf{r}}\hat{\mathbf{p}}]; \hat{\tilde{Q}}_{ij} = \frac{e}{2mc}\left(\hat{r}_i\hat{p}_j + \hat{r}_j\hat{p}_i\right) \tag{4}$$

operators of magnetic dipole and electric quadrupole for one-electron quantum system under consideration.

In first order of perturbation theory, the transition rate from an initial atomic state $|i\rangle$ to a final state $|f\rangle$ is given by Fermi's golden rule (Fermi 1932),

$$R_{fi} = \frac{2\pi}{\hbar}\sum_{I,F} p(I)\langle I|\bar{\hat{H}}_{\text{int}}^{-}(\mathbf{r}_0)|F\rangle\langle F|\bar{\hat{H}}_{\text{int}}^{+}(\mathbf{r}_0)|I\rangle \times \\ \delta\left(E_F + E_f - E_I - E_i\right) \tag{5}$$

where

$$\bar{\hat{H}}_{\text{int}} = -\hat{\mathbf{A}}(\mathbf{r}_0)\tilde{\mathbf{p}} - \hat{\mathbf{H}}(\mathbf{r}_0)\mathbf{m} - \frac{\partial \hat{A}_i(\mathbf{r}_0)}{\partial r_{0,j}}\tilde{Q}_{ij} \\ \tilde{\mathbf{p}} = \frac{e}{mc}\langle f|\hat{\mathbf{p}}|i\rangle; \mathbf{m} = \langle f|\hat{\mathbf{m}}|i\rangle; \tilde{Q}_{ij} = \langle f|\hat{\tilde{Q}}_{ij}|i\rangle \tag{6}$$

matrix element of interaction Hamiltonian between atomic states and ± signs stands for positive and negative frequency parts of field operators (Glauber1963) .

Capital letters denote eigenstates of the free electromagnetic field. Such eigenstates might involve, and depend on the coupling between, the radiation field, other atoms, surface excitations, and the like. For convenience, we refer to these as "field states". *p(I)* stands for probability to find initial field in state I. Small letters in subsriptws correspond to eigenstates of detector.

Expressing Dirac δ function in (5) in integral form we find

$$R_{fi}=\frac{1}{\hbar^2}\int_{-\infty}^{\infty}dt\exp(i\omega_0 t)$$

$$\begin{Bmatrix}\left\langle \hat{A}_i^-(\mathbf{r}_0,t)\hat{A}_j^+(\mathbf{r}_0',0)\right\rangle \tilde{p}_i\bar{p}_j+\left\langle \hat{H}_i^-(\mathbf{r}_0,t)\hat{H}_j^+(\mathbf{r}_0',0)\right\rangle m_i\bar{m}_j+\left\langle \frac{\partial\hat{A}_i^-}{\partial r_{0,j}}(\mathbf{r}_0,t)\frac{\partial\hat{A}_k^+}{\partial r_{0,l}}(\mathbf{r}_0,t)\right\rangle \tilde{Q}_{ij}\bar{\tilde{Q}}_{kl}+\\ \left\langle \hat{A}_i^-(\mathbf{r}_0,t)\hat{H}_j^+(\mathbf{r}_0',0)\right\rangle \tilde{p}_i\bar{m}_j+\left\langle \hat{H}_i^-(\mathbf{r}_0,t)\hat{A}_j^+(\mathbf{r}_0',0)\right\rangle m_i\bar{p}_j+\\ \left\langle \hat{A}_i^-(\mathbf{r}_0,t)\frac{\partial\hat{A}_k^+}{\partial r_{0,l}}(\mathbf{r}_0,t)\right\rangle \tilde{p}_i\bar{\tilde{Q}}_{kl}+\left\langle \frac{\partial\hat{A}_k^-}{\partial r_{0,l}}(\mathbf{r}_0,t)\hat{A}_j^+(\mathbf{r}_0',0)\right\rangle \tilde{Q}_{kl}\bar{p}_j+\\ \left\langle \hat{H}_i^-(\mathbf{r}_0,t)\frac{\partial\hat{A}_k^+}{\partial r_{0,l}}(\mathbf{r}_0,t)\right\rangle m_i\bar{\tilde{Q}}_{kl}+\left\langle \frac{\partial\hat{A}_i^-}{\partial r_{0,j}}(\mathbf{r}_0,t)\hat{H}_k^+(\mathbf{r}_0',0)\right\rangle \tilde{Q}_{ij}\bar{m}_k\end{Bmatrix} \tag{7}$$

where $\omega_0=(E_f-E_i)/\hbar$, the Latin subscripts denote Cartesian coordinates and are to be summed over when repeated. In eq. (7) angular brackets indicate an ensemble average $\langle ....\rangle=\sum_I p(I)\langle I|....|I\rangle$ and $\hat{A}(\mathbf{r}_0,t)$ is an interaction picture operator, evolving as if (1) were not present,

$$\hat{\mathbf{A}}(\mathbf{r},t)=\exp\left(\frac{i}{\hbar}\hat{H}_0 t\right)\hat{\mathbf{A}}(\mathbf{r},0)\exp\left(-\frac{i}{\hbar}\hat{H}_0 t\right) \tag{8}$$

In (8) $\hat{H}_0$ is Hamiltonian of the free electromagnetic field.

Thus, the excitation rate for arbitrary optical field can be expressed through gradients of different correlation functions and atomic matrix elements.

In general spatial behavior of correlation functions in (7) depends on initial conditions and should be stated by special theoretical and experimental investigations. Here for simplicity we will consider only coherent (quasi classical) initial state of optical fields. In this state correlation functions factorize, eg

$$\begin{aligned}\left\langle \hat{A}_i^-(\mathbf{r}_0,t)\hat{A}_j^+(\mathbf{r}_0',0)\right\rangle&=\left\langle \hat{A}_i^-(\mathbf{r}_0,t)\right\rangle\left\langle \hat{A}_j^+(\mathbf{r}_0',0)\right\rangle\\ \left\langle \hat{H}_i^-(\mathbf{r}_0,t)\hat{A}_j^+(\mathbf{r}_0',0)\right\rangle&=\left\langle \hat{H}_i^-(\mathbf{r}_0,t)\right\rangle\left\langle \hat{A}_j^+(\mathbf{r}_0',0)\right\rangle\end{aligned} \tag{9}$$

where $\left\langle \hat{A}_i(\mathbf{r},t)\right\rangle=A_i(\mathbf{r},t)$ is mean value of vector potential that is its classical values. Analogous situation occur for other fields. As a result, excitation rate will have the form

$$R_{i\to f}=\frac{T^{-}(\mathbf{r}_0,\omega_0)T^{+}(\mathbf{r}_0,t=0)}{\hbar^2},T^{-}(\mathbf{r}_0,\omega_0)=\int dte^{i\omega_0 t}T^{-}(\mathbf{r}_0,t)$$
$$T^{-}(\mathbf{r},\omega_0)=\mathbf{A}(\mathbf{r},\omega_0)\tilde{\mathbf{p}}+\mathbf{B}(\mathbf{r},\omega_0)\mathbf{m}+\frac{\partial A_i(\mathbf{r},\omega_0)}{\partial r_{0,j}}\tilde{Q}_{ij} \tag{10}$$
$$T^{+}(\mathbf{r},t=0)=\mathbf{A}(\mathbf{r},t=0)\tilde{\mathbf{p}}+\mathbf{B}(\mathbf{r},t=0)\mathbf{m}+\frac{\partial A_i(\mathbf{r},t=0)}{\partial r_{0,j}}\tilde{Q}_{ij}$$

For resonant excitation beam with narrow linewidth we will have instead of (10) the following expression

$$R_{i\to f}=\frac{\left|T^{if}\right|^2}{\hbar^2\Delta}$$
$$T^{if}=T^{if}_{E1}+T^{if}_{M1}+T^{if}_{E2}=\mathbf{d}^{if}\mathbf{E}(\mathbf{r},\omega_0)+\mathbf{m}^{if}\mathbf{B}(\mathbf{r},\omega_0)+Q_{ij}\nabla_i E_j(\mathbf{r},\omega_0) \tag{11}$$

where $\mathbf{d}=\frac{\tilde{\mathbf{p}}}{\omega_0};Q_{ij}=\frac{\tilde{Q}_{ij}}{\omega_0}$ matrix elements of electric dipole and quadrupole operators and $\Delta$ stands for characteristic excitation linewidth or transition linewidth, ($\Delta^2=\delta\omega^2+\Gamma^2/4$). In derivation (11) we also use $\mathbf{E}=-\frac{1}{c}\dot{\mathbf{A}}=\frac{i\omega_0}{c}\mathbf{A}$ relation, which is valid for free fields in Coulomb gauge. In what follows we also assume that the orientation of the detector (molecule) is fixed in space and no additional averaging is needed.

For quantum detector with arbitrary internal structure the matrix elements ***d***, *m, Q* are simultaneously nonzero quantities which have 3,3 and 5 independent components respectively. These matrix elements can be parameterized in the following form within Cartesian co-ordinates (x,y,z) where z is quantization axis:

$$\begin{aligned}&\mathbf{d}^{(\pm1)}=d^{(\pm1)}(\pm1,i,0);\mathbf{d}^{(0)}=d^{(0)}(0,0,\sqrt{2})\\&\mathbf{m}^{(\pm1)}=m^{(\pm1)}(\pm1,i,0);\mathbf{m}^{(0)}=m^{(0)}(0,0,\sqrt{2})\end{aligned} \tag{12}$$

$$\mathbf{Q}^{(0)}=Q^{(0)}\sqrt{\frac{2}{3}}\begin{Vmatrix}-1&0&0\\0&-1&0\\0&0&2\end{Vmatrix};\mathbf{Q}^{(\pm1)}=Q^{(\pm1)}\begin{Vmatrix}0&0&\mp1\\0&0&-i\\\mp1&-i&0\end{Vmatrix};\quad\mathbf{Q}^{(\pm2)}=Q^{(\pm2)}\begin{Vmatrix}1&\pm i&0\\\pm i&-1&0\\0&0&0\end{Vmatrix} \tag{13}$$

If one considers spherically symmetric atom, then there will be conservation of angular momentum and superscript will correspond to z- component of angular momentum of an atom. In this case we have instead of (12) and (13)

$$Q_{ij}^{M} = Q^{R}\int \sin\theta d\theta d\varphi Y_{0}^{*0}\left(3r_i r_j - r^2\delta_{ij}\right)Y_{2}^{M}\left(\theta,\varphi\right) \quad (14)$$

$$d_{i}^{M} = d^{R}\int \sin\theta d\theta d\varphi Y_{0}^{*0} r_i Y_{1}^{M}\left(\theta,\varphi\right) \quad (15)$$

$$m_{i}^{M} = m^{R}\int \sin\theta d\theta d\varphi Y_{1}^{*0}[\mathbf{r}\nabla]_i Y_{1}^{M}\left(\theta,\varphi\right) \quad (16)$$

In this case there is additional relations between different components of $d^{(M)}, m^{(M)} (M = 0, \pm 1)$ and $Q^{(M)} (M = 0, \pm 1, \pm 2)$. It means that all $d^M$ and $Q^M$ are equal, and $m^{(1)}=m^{(-1)}=m^R$ while $m^{(0)}=0$.

In the case of general nonsymmetrical quantum system all components of $d^{(M)}, m^{(M)} (M = 0, \pm 1)$ and $Q^{(M)} (M = 0, \pm 1, \pm 2)$ are independent and parameterizations (12) and (13) have no relation to angular momentum. In this case one should consider (12) and (13) as formal but useful definitions.

### *3. Properties of free propagating Laguerre-Gauss beams*

The geometry of Laguerre-Gauss beam is shown in Fig. 0. In the case of Laguerre-Gauss beams electric fields can be presented by the following formulae (Santamato 2004)

$$\mathbf{E}^{(m)}(\mathbf{r},\omega) = E_0 \frac{w_0}{k}\left\{k\alpha U^{(m)}, k\beta U^{(m)}, i\left(\alpha\frac{\partial U^{(m)}}{\partial x} + \beta\frac{\partial U^{(m)}}{\partial y}\right)\right\}e^{ikz}$$

$$U^{(m)} = \frac{C_p^{|m|}}{w(z)}\left[\frac{\sqrt{2}r}{w(z)}\right]^{|m|}\exp\left(-\frac{r^2}{w^2(z)}\right)L_p^{|m|}\left(\frac{2r^2}{w^2(z)}\right)\times \quad (17)$$

$$\exp\left(\frac{ikr^2}{2\left(z^2+z_R^2\right)} - im\varphi - i\left(2p+|m|+1\right)\arctan\left(z/z_R\right)\right)$$

where $C_p^{|m|} = \sqrt{2p!/\pi\left(p+|m|\right)!}$ is the normalization constant, $w(z) = w_0\sqrt{1+z^2/z_R^2}$ is the beam radius at z, $w_0$ is the Gaussian beam waist, $L_p^{|m|}(x)$ is the generalized Laguerre polynomial of order p and argument *x*, and $z_R = kw_0^2/2$ is the Rayleigh range of the beam, and, finally, $(2p + |m| + 1)\arctan(z/z_R)$ is the Gouy phase. $(\alpha,\beta)$ is polarization vector.

The number p + *1* is the number of nodes of the field in the radial direction and, what is more important, the number *m* is the orbital angular momentum carried by the beam along its propagation direction in units of $\hbar$. We may notice, in fact, that LG modes are proportional to $r^{|m|}e^{-im\varphi} = (x \pm iy)^{|m|}$, a term characteristic of the eigenfunctions of the orbital angular momentum operator $l_z = -i\hbar\,(x\partial_y - y\partial_x) = -i\hbar\,\partial_\varphi$ .

In the following table some lower U's are shown for z=0 (without factor $e^{-r^2/w_0^2}$)

Table 1. Expression for $U^{(m)}$ –function at the waist plane (z=0) without factor $e^{-r^2/w_0^2}$.

| | m=0 | m=1 | m=2 |
|---|---|---|---|
| p=0 | $\frac{\sqrt{2}}{\sqrt{\pi}w_0}$ | $\frac{2r}{\sqrt{\pi}w_0^2}e^{-i\varphi}$ | $\frac{2r^2}{\sqrt{\pi}w_0^3}e^{-2i\varphi}$ |
| p=1 | $-\frac{\sqrt{2}}{\sqrt{\pi}}\frac{(2r^2-w_0^2)}{w_0^3}$ | $-\frac{2r\sqrt{2}}{\sqrt{\pi}}\frac{(r^2-w_0^2)}{w_0^4}e^{-i\varphi}$ | $\frac{2\sqrt{3}}{3\sqrt{\pi}}\frac{r^2(3w_0^2-2r^2)}{w_0^5}e^{-2i\varphi}$ |
| p=2 | $\frac{\sqrt{2}}{\sqrt{\pi}}\frac{(w_0^4-4w_0^2r^2+2r^4)}{w_0^5}$ | $\frac{2\sqrt{3}}{3\sqrt{\pi}}\frac{r(3w_0^4-6w_0^2r^2+2r^4)}{w_0^6}e^{-i\varphi}$ | $\frac{2\sqrt{6}}{3\sqrt{\pi}}\frac{r^2(3w_0^4-4w_0^2r^2+r^4}{w_0^7}e^{-2i\varphi}$ |

The most important properties of Laguerre-Gauss beams is that they can bear both spin and orbital angular momentum, and total averaged momentum can be described by the formulae

$$j_z = \hbar(m+\sigma), \sigma = -i(\alpha\beta^* - \beta\alpha^*) \tag{18}$$

This formulae is derived for the case where is no interaction of beam with matter. It is important to note that eq (18) is approximate expression and for highly focused beams one should use more correct formulae instead of (18) (Barnett Allen 1994).

The electric field vector (17) has longitudinal z-component (see eg Santamatoto 2004 ). The presence of longitudinal fields is necessary to provide charge conservation law div**E**=0. Strictly speaking even (12) does not provide charge conservation law because div**E** ~ $1/k^2 \neq 0$. It is possible to correct (12) but in what follows there is no necessity to do it.

Magnetic field corresponding to (17) can be found from Faraday's law of induction $ik\mathbf{H} = rot\mathbf{E}$ and has the following form

$$H^{(m)}(\mathbf{r},\omega) = E_0 \frac{w_0}{k} \begin{Bmatrix} -k\beta U^{(m)} + i\beta \frac{\partial U^{(m)}}{\partial z} + \frac{\alpha}{k}\frac{\partial^2 U^{(m)}}{\partial x \partial y} + \frac{\beta}{k}\frac{\partial^2 U^{(m)}}{\partial y^2}, \\ k\alpha U - i\alpha \frac{\partial U^{(m)}}{\partial z} - \frac{\alpha}{k}\frac{\partial^2 U^{(m)}}{\partial x^2} - \frac{\beta}{k}\frac{\partial^2 U^{(m)}}{\partial x \partial y}, \\ i\left(\alpha \frac{\partial U^{(m)}}{\partial y} - \beta \frac{\partial U^{(m)}}{\partial x}\right) \end{Bmatrix} e^{ikz} \quad (19)$$

The key feature of nontrivial (m>1) LG beams is that all components of the electric field and the electric energy density of (17) are equal to zero at the axis (see Table 1). Due to this fact LG beams are often referred to as hollow beams and "doughnut beams".

However one should remember about magnetic energy density and gradients of electric field at the axis. For some cases magnetic energy is not zero at the axis, and some optical transitions can occur even for molecules placed at the axis of the LG beam. An important case is the case with $m = \pm 2$. In this case the following components are nonzero only for *m=2*

$$\begin{aligned} B_x(0) &= -i\frac{\sqrt{8(p+1)(p+2)}}{k^2\sqrt{\pi}w_0^2}(\alpha - i\beta)E_0 \\ B_y(0) &= -\frac{\sqrt{8(p+1)(p+2)}}{k^2\sqrt{\pi}w_0^2}(\alpha - i\beta)E_0 \\ \frac{\partial E_z}{\partial x}(0) &= i\frac{\sqrt{8(p+1)(p+2)}}{\sqrt{\pi}w_0^2 k}(\alpha - i\beta)E_0 \\ \frac{\partial E_z}{\partial y}(0) &= \frac{\sqrt{8(p+1)(p+2)}}{\sqrt{\pi}w_0^2 k}(\alpha - i\beta)E_0 \end{aligned} \quad (20)$$

while for m=-2 we have

$$
\begin{aligned}
B_x(0) &= i\frac{\sqrt{8(p+1)(p+2)}}{k^2\sqrt{\pi}w_0^2}(\alpha+i\beta)E_0 \\
B_y(0) &= -\frac{\sqrt{8(p+1)(p+2)}}{k^2\sqrt{\pi}w_0^2}(\alpha+i\beta)E_0 \\
\frac{\partial E_z}{\partial x}(0) &= i\frac{\sqrt{8(p+1)(p+2)}}{\sqrt{\pi}w_0^2 k}(\alpha+i\beta)E_0 \\
\frac{\partial E_z}{\partial y}(0) &= -\frac{\sqrt{8(p+1)(p+2)}}{\sqrt{\pi}w_0^2 k}(\alpha+i\beta)E_0
\end{aligned}
\qquad (21)
$$

Other components are equal to zero.

From these equations it is easy to find that in this case the electric energy density is zero on the axis, while magnetic energy density, $I_M$, is nonzero:

$$
I_M = \frac{1}{16\pi}|B|^2 = \frac{E_0^2}{16\pi}\frac{16(p+1)(p+2)}{\pi(kw_0)^4}|\alpha \mp i\beta|^2 \qquad (22)
$$

Below (Fig.2,3) for distinctness we will consider LG beams with $\alpha = 1/\sqrt{2}, \beta = i/\sqrt{2}$ case. It corresponds to LG beam with spin equal to -1. So the total angular momentum of our LG beam will be described by $j_z = \hbar(m-1)$.The dependence of electric and magnetic energy on radius is shown in Fig. 2. From this figure one can see that magnetic energy on axis is comparable even with electric energy density at its maximum. In Fig.3 the ratio of magnetic energy at the axis to electric energy density at its maximum is shown. One notices that substantial magnetic energy density occurs for strongly focused beams.

From formal point of view nonzero radial magnetic fields on the beam axis are due to presence of longitudinal electric fields in the beam. These longitudinal electric fields are more pronounced for more focused beam and for more zeros in radial direction. In its turn longitudinal electric fields due to Faraday's law of electromagnetic induction results in nonzero radial magnetic fields on the axis. More deep insight in the problem shows that nonzero magnetic ( or electric) fields on the axis are related with difficulties in definition of unique phase of vector fields (Bialynicki-Birula Bialynicki-Birula 2003).

The spatial structure of Laguerre-Gauss beam is very complicated in comparison with usual circular polarized light. In Fig. 4-7 the distribution of electric and magnetic field at the waist plane is shown. for the case of circular polarization ($\alpha = 1/\sqrt{2}, \beta = i/\sqrt{2}$) while in Fig 8-11 such distributions are plotted for linear polarization $\alpha = 1, \beta = 0$ From these figures the complicated magnetic structure of LG beams becomes evident. The most interesting feature is that the magnetic field and gradients of electric fields are nonzero at axis and Glauber's ideal photon detector cannot work here.

Another interesting characteristic is that even in the case of linear polarization of electric fields, magnetic fields on the axis will rotate in space with optical frequency.

**4.Laguerre-Gauss beam: interaction with elementary quantum system**

As we already have seen although electric fields are zero on the axis, the magnetic fields $B_x, B_y$ and gradients of electric fields $\frac{\partial E_z}{\partial x}, \frac{\partial E_z}{\partial y}$ of LG beam with m=2 are nonzero on the axis. Obviously very interesting effects can occur due this portion of electromagnetic energy. To test these effects we suggest using new type of detectors described above.

Below we will consider interaction of Laguerre-Gauss beams bearing different angular momenta m =0,±1,±2 with detector of general structure (see eq. (12)(13) for its matrix elements).

The results of calculation of matrix elements for excitation rates for different types of detector placed at axis of LG beams are presented below (Tables 2-13).

Table 2. Values of excitation amplitudes $T_{E1}^{mM}$ on the axis (x=y=z=0) (arbitrary polarization)

| | M=-1 | M=0 | M=1 |
|---|---|---|---|
| m=-2 | 0 | 0 | 0 |
| m=-1 | 0 | $-\frac{2\sqrt{2}\sqrt{p+1}}{\sqrt{\pi}kw_0}(\alpha+i\beta)E_0d^0$ | 0 |
| m=0 | $\frac{-i\sqrt{2}(\alpha-i\beta)}{\sqrt{\pi}}E_0d^{(-1)}$ | 0 | $\frac{i\sqrt{2}(\alpha+i\beta)}{\sqrt{\pi}}E_0d^{(1)}$ |
| m=1 | 0 | $-\frac{2\sqrt{2}\sqrt{p+1}}{\sqrt{\pi}kw_0}(\alpha-i\beta)E_0d^0$ | 0 |
| m=2 | 0 | 0 | 0 |

Table 3. Values of excitation amplitudes $T_{M1}^{mM}$ on the axis (x=y=z=0)

(arbitrary polarization)

| | M=-1 | M=0 | M=1 |
|---|---|---|---|
| m=-2 | $\frac{4\sqrt{2}\sqrt{(p+1)(p+2)}}{\sqrt{\pi}w_0^2k^2}(\alpha+i\beta)E_0m^{(1)}$ | 0 | 0 |
| m=-1 | 0 | - $\frac{2\sqrt{2}i\sqrt{p+1}}{\sqrt{\pi}kw_0}(\alpha+i\beta)E_0m^{(0)}$ | 0 |
| m=0 | $\frac{-\sqrt{2}}{\sqrt{\pi}}(\alpha-i\beta)E_0m^{(-1)}$ | 0 | $\frac{-\sqrt{2}}{\sqrt{\pi}}(\alpha+i\beta)E_0m^{(-1)}$ 0 |
| m=1 | 0 | $\frac{2\sqrt{2}i\sqrt{p+1}}{\sqrt{\pi}kw_0}(\alpha-i\beta)E_0m^{(0)}$ | 0 |
| m=2 | 0 | 0 | $\frac{4\sqrt{2}\sqrt{(p+1)(p+2)}}{\sqrt{\pi}w_0^2k^2}(\alpha-i\beta)E_0m^{(1)}$ |

Table 4. Values of excitation amplitudes $T_{E2}^{mM}$ on the axis (x=y=z=0)

(arbitrary polarization)

| | M=-2 | M=-1 | M=0 | M=1 | M=2 |
|---|---|---|---|---|---|
| -2 | 0 | $-\frac{4\sqrt{2(p+1)(p+2)}}{\sqrt{\pi}kw_0^2}$ $\times(\alpha+i\beta)E_0Q^R$ | 0 | 0 | 0 |
| -1 | $\frac{4i\sqrt{p+1}}{\sqrt{\pi}w_0}(\alpha-i\beta)$ | 0 | $i\frac{2\sqrt{2(p+1)}}{\sqrt{3}\sqrt{\pi}}\frac{\left(8p+8-3(kw_0)^2\right)}{k^2w_0^3}$ $\times(\alpha+i\beta)E_0Q^R$ | 0 | 0 |
| 0 | 0 | $\frac{\sqrt{2}\left(8p+4-(kw_0)^2\right)}{kw_0^2\sqrt{\pi}}$ $\times(\alpha-i\beta)E_0Q^R$ | 0 | $-\frac{\sqrt{2}\left(8p+4-(kw_0)^2\right)}{kw_0^2\sqrt{\pi}}$ $\times(\alpha+i\beta)E_0Q^R$ | 0 |
| 1 | 0 | 0 | $i\frac{2\sqrt{2}\sqrt{p+1}}{\sqrt{3}\sqrt{\pi}}\frac{\left(8p+8-3(kw_0)^2\right)}{k^2w_0^3}$ $\times(\alpha-i\beta)E_0Q^R$ | 0 | $\frac{4i\sqrt{(p+1)}}{\sqrt{\pi}w_0}(\alpha+i\beta)E_0Q^R$ |
| 2 | 0 | 0 | 0 | $\frac{4\sqrt{2(p+1)(p+2)}}{\sqrt{\pi}kw_0^2}$ $\times(\alpha-i\beta)E_0Q^R$ | 0 |

For specific polarizations these tables of matrix elements can be substantially simplified

Table 5. Values of excitation amplitudes $T_{E1}^{mM}$ on the axis (x=y=z=0)

( circular polarization σ=-1)

| | M=-1 | M=0 | M=1 |
|---|---|---|---|
| m=-2 | 0 | 0 | 0 |
| m=-1 | 0 | 0 | 0 |
| m=0 | $\frac{-2i}{\sqrt{\pi}} E_0 d^{(-1)}$ | 0 | 0 |
| m=1 | 0 | $-\frac{4\sqrt{p+1}}{\sqrt{\pi} k w_0} E_0 d^0$ | 0 |
| m=2 | 0 | 0 | 0 |

Table 6. Values of excitation amplitudes $T_{M1}^{mM}$ on the axis (x=y=z=0) (circular polarization σ=-1)

| | M=-1 | M=0 | M=1 |
|---|---|---|---|
| m=-2 | 0 | 0 | 0 |
| m=-1 | 0 | 0 | 0 |
| m=0 | $\frac{-2}{\sqrt{\pi}}E_0 m^{(-1)}$ | 0 | 0 |
| m=1 | 0 | $\frac{4i\sqrt{p+1}}{\sqrt{\pi}kw_0}E_0 m^{(0)}$ | 0 |
| m=2 | 0 | 0 | $\frac{8\sqrt{(p+1)(p+2)}}{\sqrt{\pi}w_0^2k^2}E_0 m^{(1)}$ |

Table 7. Values of excitation amplitudes $T_{E2}^{mM}$ on the axis (x=y=z=0)

(circular polarization σ=-1)

| | M=-2 | M=-1 | M=0 | M=1 | M=2 |
|---|---|---|---|---|---|
| -2 | 0 | 0 | 0 | 0 | 0 |
| -1 | $\frac{4i\sqrt{2(p+1)}}{\sqrt{\pi}w_0}E_0Q^R$ | 0 | 0 | 0 | 0 |
| 0 | 0 | $\frac{2\left(8p+4-(kw_0)^2\right)}{kw_0^2\sqrt{\pi}}E_0Q^R$ | 0 | 0 | 0 |
| 1 | 0 | 0 | $i\frac{4\sqrt{p+1}}{\sqrt{3}\sqrt{\pi}}\frac{\left(8p+8-3(kw_0)^2\right)}{k^2w_0^3}E_0Q^R$ | 0 | 0 |
| 2 | 0 | 0 | 0 | $\frac{8\sqrt{(p+1)(p+2)}}{\sqrt{\pi}kw_0^2}E_0Q^R$ | 0 |

Table 8. Values of excitation amplitudes $T_{E1}^{mM}$ on the axis (x=y=z=0)

(circular polarization σ=1)

| | M=-1 | M=0 | M=1 |
|---|---|---|---|
| m=-2 | 0 | 0 | 0 |
| m=-1 | 0 | $-\frac{4\sqrt{p+1}}{\sqrt{\pi}kw_0}E_0d^0$ | 0 |
| m=0 | 0 | 0 | $\frac{2i}{\sqrt{\pi}}E_0d^{(1)}$ |
| m=1 | 0 | 0 | 0 |
| m=2 | 0 | 0 | 0 |

Table 9. Values of excitation amplitudes $T_{M1}^{mM}$ on the axis (x=y=z=0) (circular polarization σ=1)

| | M=-1 | M=0 | M=1 |
|---|---|---|---|
| m=-2 | $\frac{8\sqrt{(p+1)(p+2)}}{\sqrt{\pi}w_0^2k^2}E_0m^{(1)}$ | 0 | 0 |
| m=-1 | 0 | $-\frac{4i\sqrt{p+1}}{\sqrt{\pi}kw_0}E_0m^{(0)}$ | 0 |
| m=0 | 0 | 0 | $\frac{-2}{\sqrt{\pi}}E_0m^{(-1)}0$ |
| m=1 | 0 | 0 | 0 |
| m=2 | 0 | 0 | 0 |

Table 10. Values of excitation amplitudes $T_{E2}^{mM}$ on the axis (x=y=z=0)

(circular polarization σ=1)

| | M=-2 | M=-1 | M=0 | M=1 | M=2 |
|---|---|---|---|---|---|
| -2 | 0 | $-\frac{8\sqrt{(p+1)(p+2)}}{\sqrt{\pi}kw_0^2}E_0Q^R$ | 0 | 0 | 0 |
| -1 | 0 | 0 | $i\frac{4\sqrt{(p+1)}}{\sqrt{3}\sqrt{\pi}}\frac{\left(8p+8-3(kw_0)^2\right)}{k^2w_0^3}E_0Q^R$ | 0 | 0 |
| 0 | 0 | 0 | 0 | $-\frac{2\left(8p+4-(kw_0)^2\right)}{kw_0^2\sqrt{\pi}}E_0Q^R$ | 0 |
| 1 | 0 | 0 | 0 | 0 | $\frac{4i\sqrt{2(p+1)}}{\sqrt{\pi}w_0}E_0Q^R$ |
| 2 | 0 | 0 | 0 | 0 | 0 |

Table 11. Values of excitation amplitudes $T_{E1}^{mM}$ on the axis (x=y=z=0)

( linear polarization α=1 ,β=0)

| | M=-1 | M=0 | M=1 |
|---|---|---|---|
| m=-2 | 0 | 0 | 0 |
| m=-1 | 0 | $-\frac{2\sqrt{2}\sqrt{p+1}}{\sqrt{\pi}kw_0}E_0 d^0$ | 0 |
| m=0 | $\frac{-i\sqrt{2}}{\sqrt{\pi}}E_0 d^{(-1)}$ | 0 | $\frac{i\sqrt{2}}{\sqrt{\pi}}E_0 d^{(1)}$ |
| m=1 | 0 | $-\frac{2\sqrt{2}\sqrt{p+1}}{\sqrt{\pi}kw_0}E_0 d^0$ | 0 |
| m=2 | 0 | 0 | 0 |

Table 12. Values of excitation amplitudes $T_{M1}^{mM}$ on the axis (x=y=z=0)

( linear polarization α=1 ,β=0)

| | M=-1 | M=0 | M=1 |
|---|---|---|---|
| m=-2 | $\frac{4\sqrt{2}\sqrt{(p+1)(p+2)}}{\sqrt{\pi}w_0^2k^2}E_0m^{(1)}$ | 0 | 0 |
| m=-1 | 0 | $-\frac{2\sqrt{2}i\sqrt{p+1}}{\sqrt{\pi}kw_0}E_0m^{(0)}$ | 0 |
| m=0 | $\frac{-\sqrt{2}}{\sqrt{\pi}}E_0m^{(-1)}$ | 0 | $\frac{-\sqrt{2}}{\sqrt{\pi}}E_0m^{(-1)}0$ |
| m=1 | 0 | $\frac{2\sqrt{2}i\sqrt{p+1}}{\sqrt{\pi}kw_0}E_0m^{(0)}$ | 0 |
| m=2 | 0 | 0 | $\frac{4\sqrt{2}\sqrt{(p+1)(p+2)}}{\sqrt{\pi}w_0^2k^2}E_0m^{(1)}$ |

Table 13. Values of excitation amplitudes $T_{E2}^{mM}$ on the axis (x=y=z=0)

( linear polarization α=1 ,β=0)

| | M=-2 | M=-1 | M=0 | M=1 | M=2 |
|---|---|---|---|---|---|
| -2 | 0 | $-\frac{4\sqrt{2(p+1)(p+2)}}{\sqrt{\pi}kw_0^2}E_0Q^R$ | 0 | 0 | 0 |
| -1 | $\frac{4i\sqrt{p+1}}{\sqrt{\pi}w_0}E_0Q^R$ | 0 | $i\frac{2\sqrt{2(p+1)}}{\sqrt{3}\sqrt{\pi}}\frac{\left(8p+8-3(kw_0)^2\right)}{k^2w_0^3}E_0Q^R$ | 0 | 0 |
| 0 | 0 | $\frac{\sqrt{2}\left(8p+4-(kw_0)^2\right)}{kw_0^2\sqrt{\pi}}(\alpha -$ | 0 | $-\frac{\sqrt{2}\left(8p+4-(kw_0)^2\right)}{kw_0^2\sqrt{\pi}}E_0Q^R$ | 0 |
| 1 | 0 | 0 | $i\frac{2\sqrt{2}\sqrt{p+1}}{\sqrt{3}\sqrt{\pi}}\frac{\left(8p+8-3(kw_0)^2\right)}{k^2w_0^3}E_0Q^R$ | 0 | $\frac{4i\sqrt{(p+1)}}{\sqrt{\pi}w_0}E_0Q^R$ |
| 2 | 0 | 0 | 0 | $\frac{4\sqrt{2(p+1)(p+2)}}{\sqrt{\pi}kw_0^2}E_0Q^R$ | 0 |

The analysis of these tables shows that, first of all, there is no direct ( that is for m=M) interaction between orbital moments of light beam and detecting quantun system. From this fact it immediately follows that beam bears the spin momentum. The presence of spin becomes especially evident in the case of circular polarizations, that is for $\alpha = 1/\sqrt{2}, \beta = \pm i/\sqrt{2}$.In this case matrix elements $\mathbf{m}^{(M)}\mathbf{B}^{(m)}, \mathbf{d}^{(M)}\mathbf{E}^{(m)}, Q^{M}\nabla\mathbf{E}^{(m)}$ are nonzero only for $m \pm 1 = M$, that is if they lies over main diagonal ($\sigma = 1,$ )or below it($\sigma = -1$). From these results one can arrive to the conclusion that such a beam has z-component of angular momentum $j_z = \hbar(m+\sigma)$ where $\sigma = \pm 1,$ is spin of beam, what is in agreement with independent approximate calculations of angular momentum of the beam (18) and conservation of angular momentum.

From other hand, in the case of linear polarization $(\alpha = 1, \beta = 0)$ eq. (18) gives zero value of spin. At first glance this fact should result in appearing of diagonal terms in tables of $T^{mM}$ amplitudes. However, as we know (see Tables 11,12,13) it is not true and linear polarization should be considered as superposition of 2 circular polarizations.

Besides one should note that in the case of E2 transitions in an atom total angular momentum of atom increases from $0\hbar$ to $2\hbar$. It is interesting that transfer of $2\hbar$ of z-component of *angular* momentum of beam to $2\hbar$ of z-component of *orbital* momentum of atom is possible through $T^{-1,-2}$ amplitude for $\sigma = -1$ or $T^{1,2}$ amplitude for $\sigma = 1,$

The most interesting is interaction of m=2 LG beam with M=1 transitions in molecules situated on the axis. First of all one can see that at the axis there is no interaction with E1 transition, while there is effective interaction with E2 and M1 transitions. The excitation rate for such transition can be written in the form:

$$R = E_0^2 \frac{64(p+1)(p+2)}{\hbar^2 \Delta\pi (kw_0)^4} \left| m^{(1)} + kQ^{(1)} \right|^2 \tag{23}$$

where $m^{(1)}$ and $Q^{(1)}$ are values of magnetic dipole and electric quadrupole matrix elements ( see (12,13)). As a rule, one can estimate $m^{(1)} \sim kQ^{(1)} \sim kea^2$, where a - characteristic size (length of DNA strand) of chiral molecules used in detector. For tuned molecules both transition can be added with the same phase which will give additional enhancement. So estimations of (23) will looks like

$$R = E_0^2 \frac{256(p+1)(p+2)}{\hbar^2 \Delta \pi (kw_0)^2} \left( \frac{a}{w_0} \right)^2 e^2 a^2 \tag{24}$$

Besides as excitation rate increases quadratically with p, one should use LG beams with high *p* to increase excitation of molecules and subsequent re-radiation of light energy. For example for $kw_0=10$, $a/w_0=0.1$ and $p=7$, excitation rates will be of the order of electric dipole excitation on plane wave ( $R_{E1} \propto \frac{E_0^2}{\hbar^2 \Delta} d^2 \propto \frac{E_0^2}{\hbar^2 \Delta} e^2 a^2$ ).

Up to now we have discussed interaction of LG beams with molecules at its axis. However, the spatial distribution of these excitation rates is also of great interest. In Fig.12-14 the dependences of excitation rate of different multipole transitions on radius are shown for LG beams with m=2 and σ=-1. From Fig12 one can see, that E2 (M=1) excitation rate has well pronounced maximum at the beam axis. This behavior should allow a spatially selective excitation of atoms or molecules located near the axis, with a sub-wavelength resolution reminiscent of confocal microscopy. In the same spirit of spatially selective excitation, a fascinating possibility apparently offered with LG beams is the selective manipulation of chiral molecules (the optical activity is usually associated to a coupled E1-M1 transition) deposited somewhere close to the hollow region. Although a negative experimental result was obtained in (Araoka et al 2005) our investigation of the local properties of the electromagnetic field suggests that it may be a too large spatial averaging that has made a possible effect unobservable. For Raman scattering of LG beams by atoms and molecules this selectivity will be even more pronounced because scattering cross-section is proportional to 4th power of field gradients.

On the other hand density of electromagnetic energy has minimum at the axis while the excitation rate for E2(M=-1) transition is equal even to zero there. Aside from the axis the situation radically changes and the excitation rate for E2(M=1) transition rapidly decreases when distance from axis increases. On the contrary the excitation rate for E2(M=-1) transition increases with distance from axis in contrary to decreasing of energy density.

Analogous situation occurs for M1 transitions. The M1(M=1) excitation rate has maximum on the axis of the LG beam. Other rates (M1(M=0,-1) are equal to zero on the axis. Aside from axis all M1 excitation rates decreases when distance from axis increases.

For E1(M=0,±1) transitions the situation is different and all partial rates (M=0,±1) are equal to zero on the axis *( r=0).*This effect is result of hollow of E field beams.

One should note that when plotting Fig12-14 we have normalized all curves as it is indicated in figures captions. However in the case of usual atoms due to selection rules some matrix elements are equal to zero ( $m^{\Delta M=0}=0$ for example). In such cases corresponding curves have obviously no sense.

## 5. Discussion

In conclusion, we suggest to use new type of detectors, which are sensitive to gradients of electric fields and to magnetic fields, in order to detect photons with a complicated space structure especially at point where electric field is zero.

We have applied our idea to LG beams for which we have shown that spiral beams should not be considered as hollow, because of nonzero magnetic fields on the axis of the beam for orbital momentum number m=±2.

One should note that not only LG beams are not hollow ones. All quasiaxially symmetric fields with m=2 should have non zero magnetic fields and gradient of electric fields at the axis. As one more examples let us now consider Bessel beams, which are exact solution of Maxwell equation. Barnett and Allen (1994) have used the fact that the Helmholtz equation is separable in cylindrical coordinates $(r,\varphi,z)$ to construct a light beam in which the electric field has the form

$$\mathbf{E}(r,\varphi,z)=\int_0^k dgf(g)\begin{Bmatrix}(\alpha\hat{\mathbf{x}}+\beta\hat{\mathbf{y}})F_m \\ +\dfrac{g}{2h}\hat{\mathbf{z}}\left[(i\alpha-\beta)F_{m+1}-(i\alpha+\beta)F_{m-1}\right]\end{Bmatrix} \tag{25}$$

Here $h=\sqrt{k^2-g^2}$ and $F_m(r,\varphi,z)=J_m(gr)\exp(i(hz-m\varphi))$

For example, for m=2 and circular polarization $\alpha=1/\sqrt{2},\beta=i/\sqrt{2}$ energy density of magnetic field at the axis will be also nonzero (Klimov et al 2008)

$$I_M=\frac{|B|^2}{8\pi}=\frac{1}{32\pi}\left|\int_0^k dgf(g)\frac{g^2}{hk}\right|^2 \tag{26}$$

while the energy of electric field still equal to zero. This fact confirms our general statement that "hollow" beams are not "hollow" in fact. On the other hand Bessel beams in waveguides have some interesting peculiarities, which will be presented in separate publication (Klimov et al 2008)

Our direct calculation of excitation rates of such transitions confirm that Laguerre-Gauss beams bear both orbital and spin momentum, which can be transferred in an elementary exchange with a quantum system, hence relaxing the usual selection rules. Despite we consider coherent state of exciting fields our conclusions and proposals are valid for any quantum state of exciting field.

From these results obtained for LG beams, we can discuss now on general grounds our suggestion of using specific detectors, which are sensitive to gradients of electric fields and to magnetic fields, in order to detect photons with a complicated space structure. The utility of gradient detectors has been proved already in hydroacoustics (Tschurov Tschurov 2002, Smaryshev 2005), where combined receivers, i.e. devices consisting of scalar sound pressure sensors and several velocity receivers (with mutually perpendicular axes) are widely used to increase sonar antennas efficiency. Our detection scheme clearly offers the possibility of detecting processes that have been intrinsically neglected in all quantum optics approaches relying only on E1-type detector. For optical experiments relying specifically on an E2 or M1 transition, an issue is certainly the low oscillator strength of these transitions, usually considered as nearly forbidden transitions Note however that the feasibility of detecting an E2 transition has already been established (Tojo Hasuo Fujimoto 2004, Tojo Hasuo 2005, Tojo Fujimoto Hasuo 2005), with an evanescent wave, that is precisely another type of e.m. field with a complicated structure (Lee et al 2006)

We also establish, through our direct calculation of excitation rates, that the high order angular momentum of LG beam can be transferred in an elementary exchange with a quantum system, hence relaxing the usual (E1) selection rules. If particle physics considerations have shown long ago that electromagnetic fields can bear a large angular momentum, which is involved in high-order multipole transition ( like in nuclear physics (Dicke 1955), the coherent production of large number of identical spiral photons is a recent achievement, up to now limited to the optical domain, but susceptible to open new frontiers in quantum optics (e.g. quantum limits to spatial correlation,...), and to encompass now a domain spanning from radio-waves to X- rays (Thide 2007 , Sasaki McNulty 2008) From a quantum optics point of view, the investigation of the intimate nature of spiral photons such as generated with LG beams has remained until now extremely limited, although the only experiment performed to date (Mair Vaziri Welhs Zeilinger 2001) at the quantum level having attracted much attention due to the opening of new sets of variable in entanglement. In particular, it should be noted that all experimental investigations involving LG beams and their specific angular momentum (Allen, Padgett Babiker 1999, Allen Barnett, Padgett 2003, Allen et al 1992, Mair Vaziri Welhs Zeilinger 2001, Tabosa, Petrov 1999, Barreiro , Tabosa 2003, Andersen 2006) have been integrated on at least a micron-size volume, instead of using a negligible size detector. Despite we consider here coherent state of exciting fields, one should note that our results apply for any quantum state of exciting field.

At last, an interesting result of our semiclassical derivation is that the specificity of the photons carried by a LG or a singular beam, appears enhanced under a strong focusing. This regime of sharply focused propagating beams opens a natural connection with the blossoming domain of nano-optics, where on the one hand, it is known that the relative strength of E2 transition might be enhanced (Klimov Ducloy 2000,2005, Klimov Letokhov 1996) but where on the other hand, quantized photons are actually attached to a material interface, instead of simply originating from a freely propagating field. With the development of nanotechnologies, it becomes conceivable to produce suitable non-E1 detectors, such as an artificial nanoparticle of special shape (nanoantennas) designed to be sensitive to gradients of electric fields (Greffet 2005). In particular, because the lower sensitivity of E2 or M1 transition is actually due to the small size of electronic orbit relatively to optical wavelength building up more sensitive non-E1 detector could be feasible with long molecules, such as twisted or bio-molecules. Besides, these detectors should benefit to the very contemporary need of characterization of effects related with very complicated electric fields between nanoparticles. For example, under some conditions the electric field intensity is equal to zero in the gap between nanoparticles (Klimov Guzatov 2007) and one should use above described detectors to characterize such field.

**Acknowledgements**

VK is grateful to the Russian Foundation for Basic Research (grants 05-02-19647, 07-02-01328) for partial financial support of this work and University Paris13 for hospitality. DB MD and RL thank French Brazilian CAPES-COFECUB (#456/04) cooperation support.

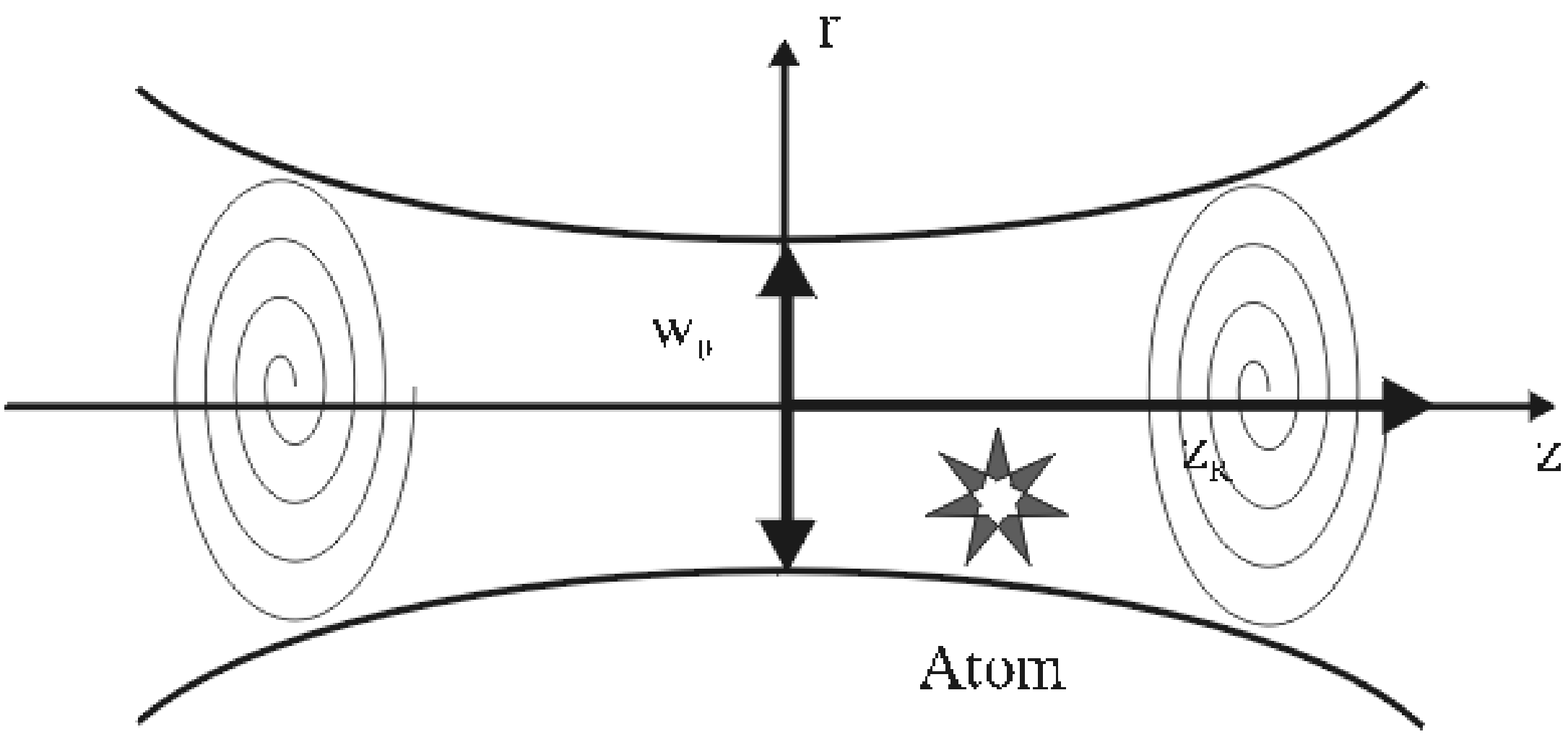

Fig.1. Geometry of Laguerre-Gauss beams

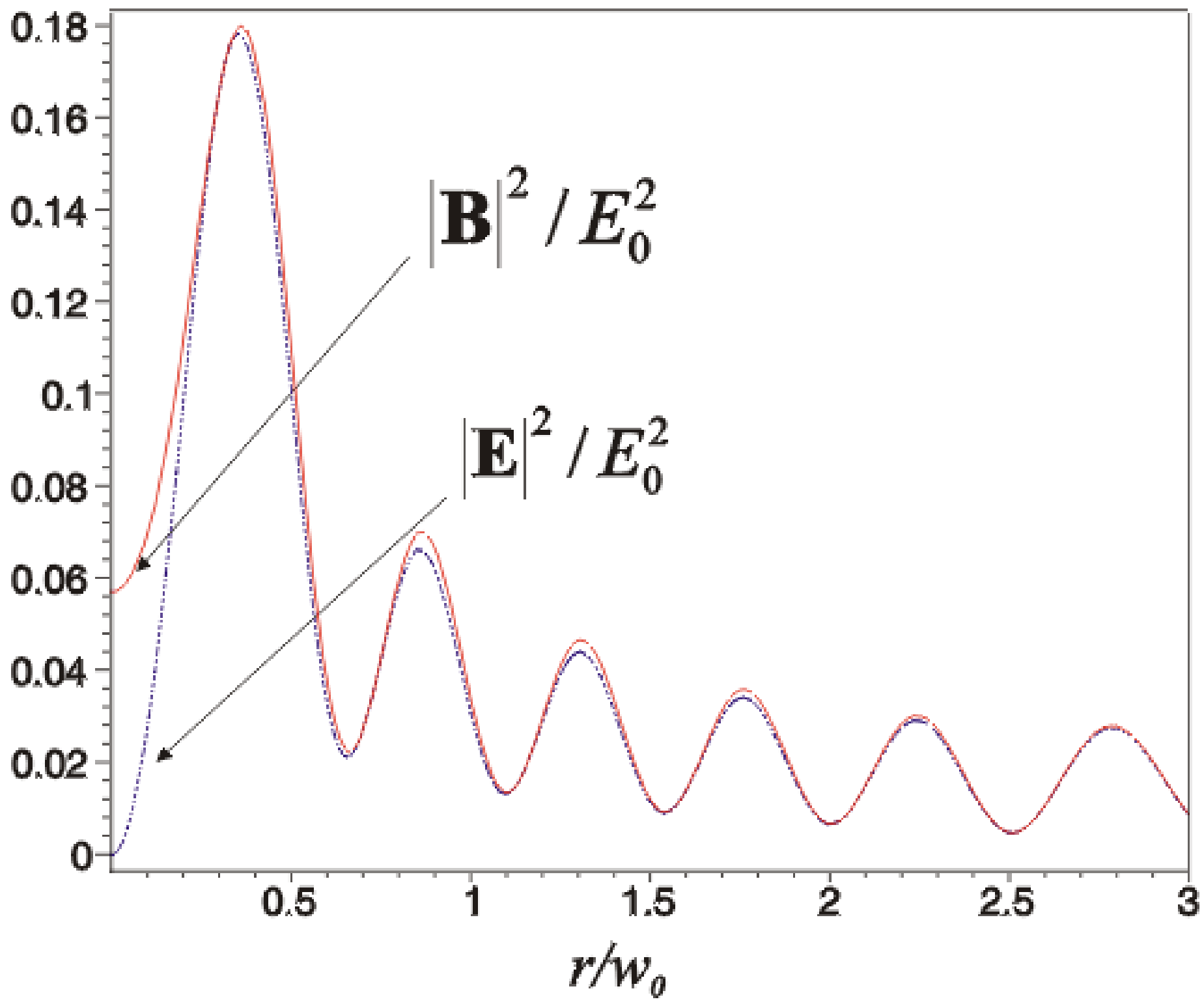

Fig.2. Dependence of electric and magnetic energy density of LG beam on radius ( $kw_0$=10,p=6,m=2, $\alpha = 1/\sqrt{2}, \beta = i/\sqrt{2}$ )

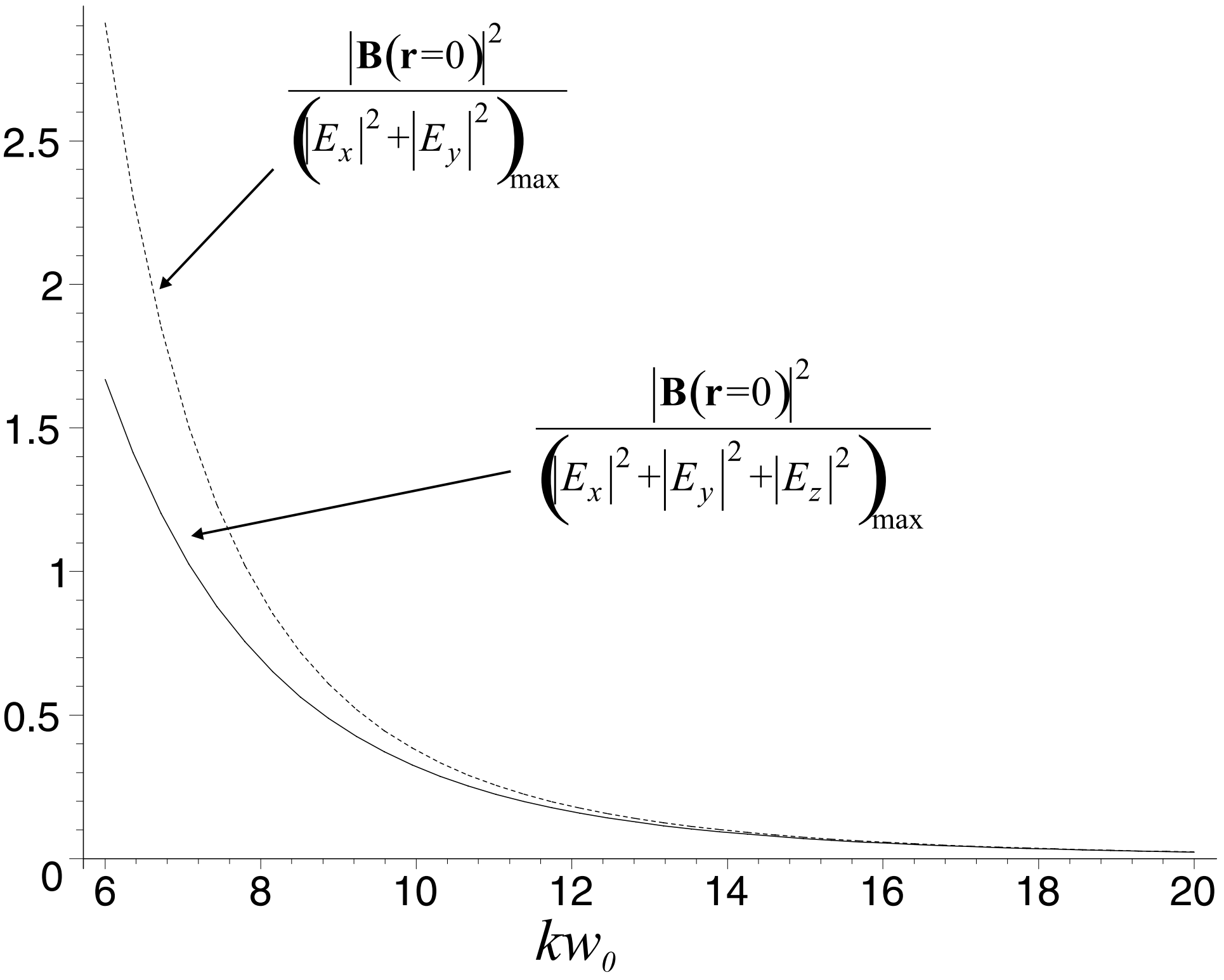

Fig.3.Ratio of magnetic energy density at center of LG beam to electric energy density at its maximum as function of beam waist $kw_0$ (p=6,m=2, $\alpha = 1/\sqrt{2}, \beta = i/\sqrt{2}$ ).

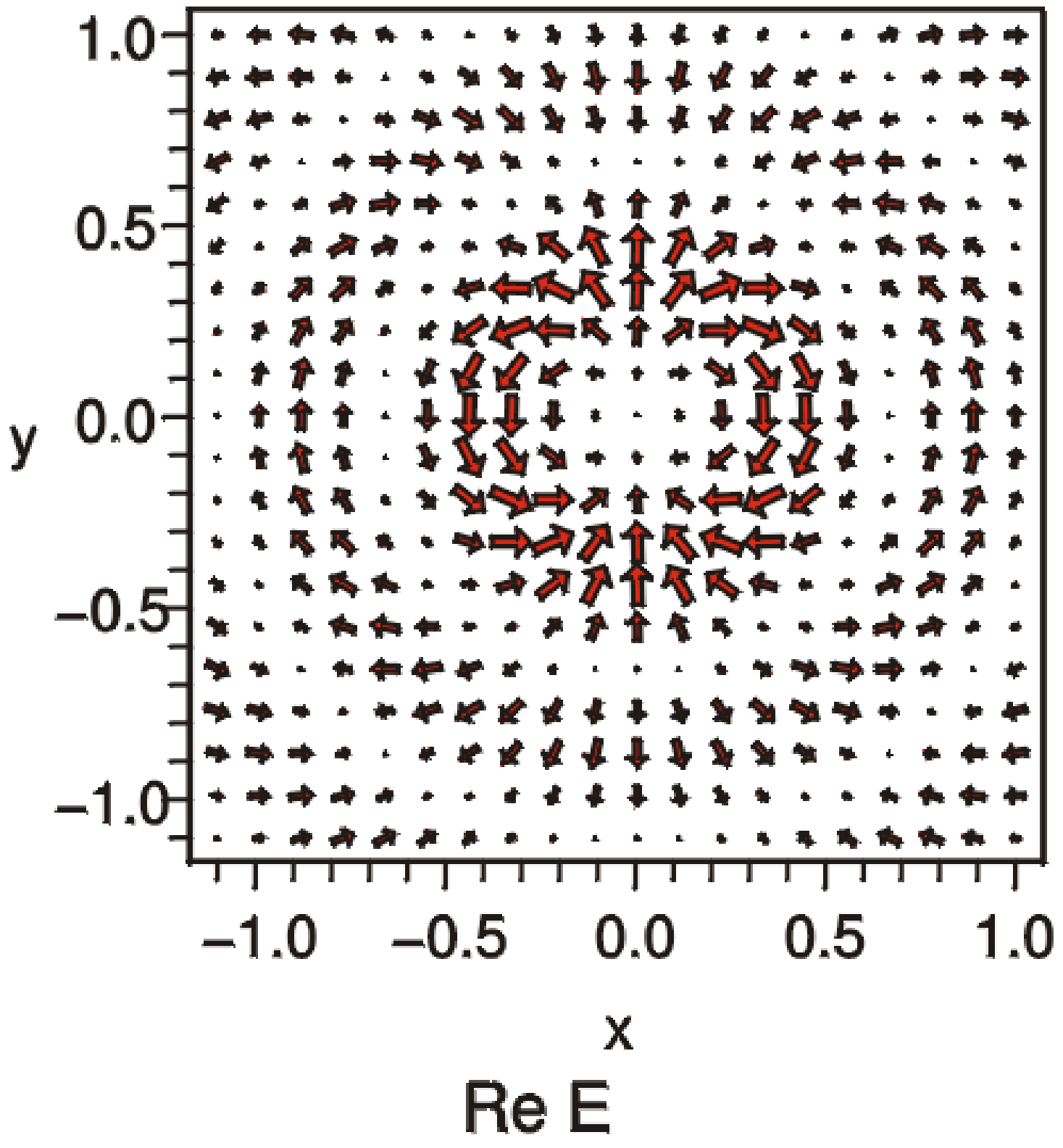

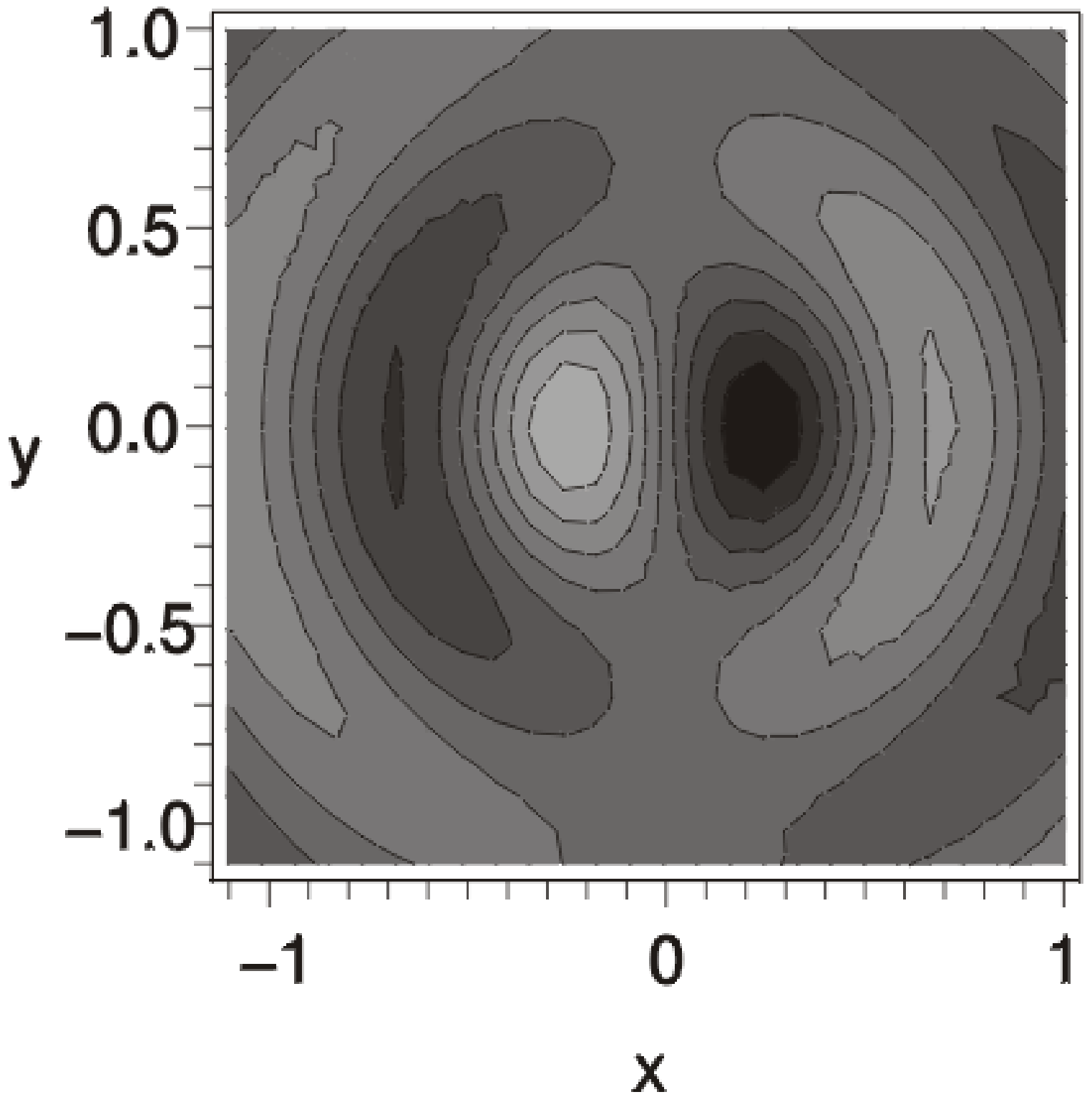

Fig.4.Spatial distribution of real part of electric field in Laguerre-Gauss beams ( $kw_0$=6,p=6,m=2,circular polarization). a) x-y components, b) z- component

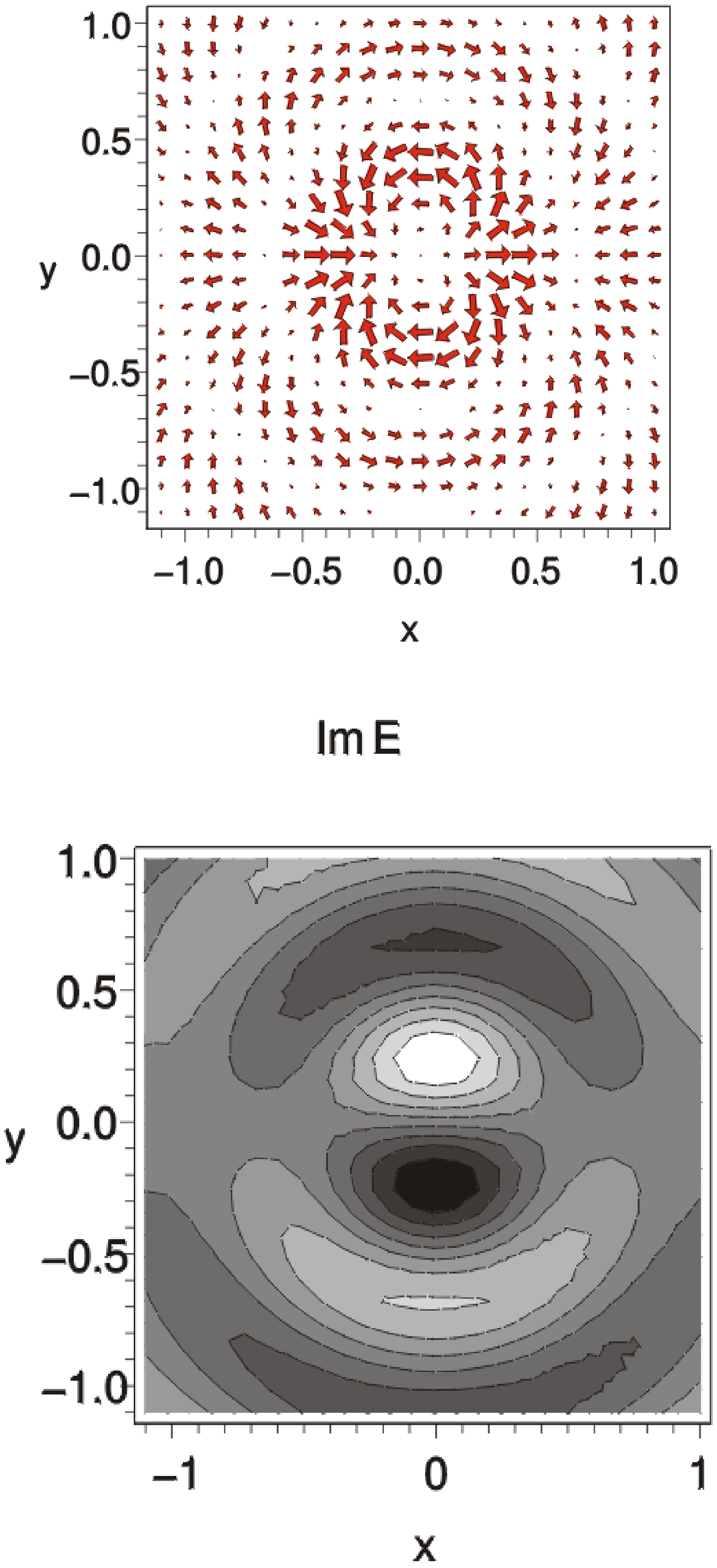

Fig.5.Spatial distribution of imaginary part of electric field in Laguerre-Gauss beams ( $kw_0$=6,p=6,m=2 circular polarization) a) x-y components, b) z- component

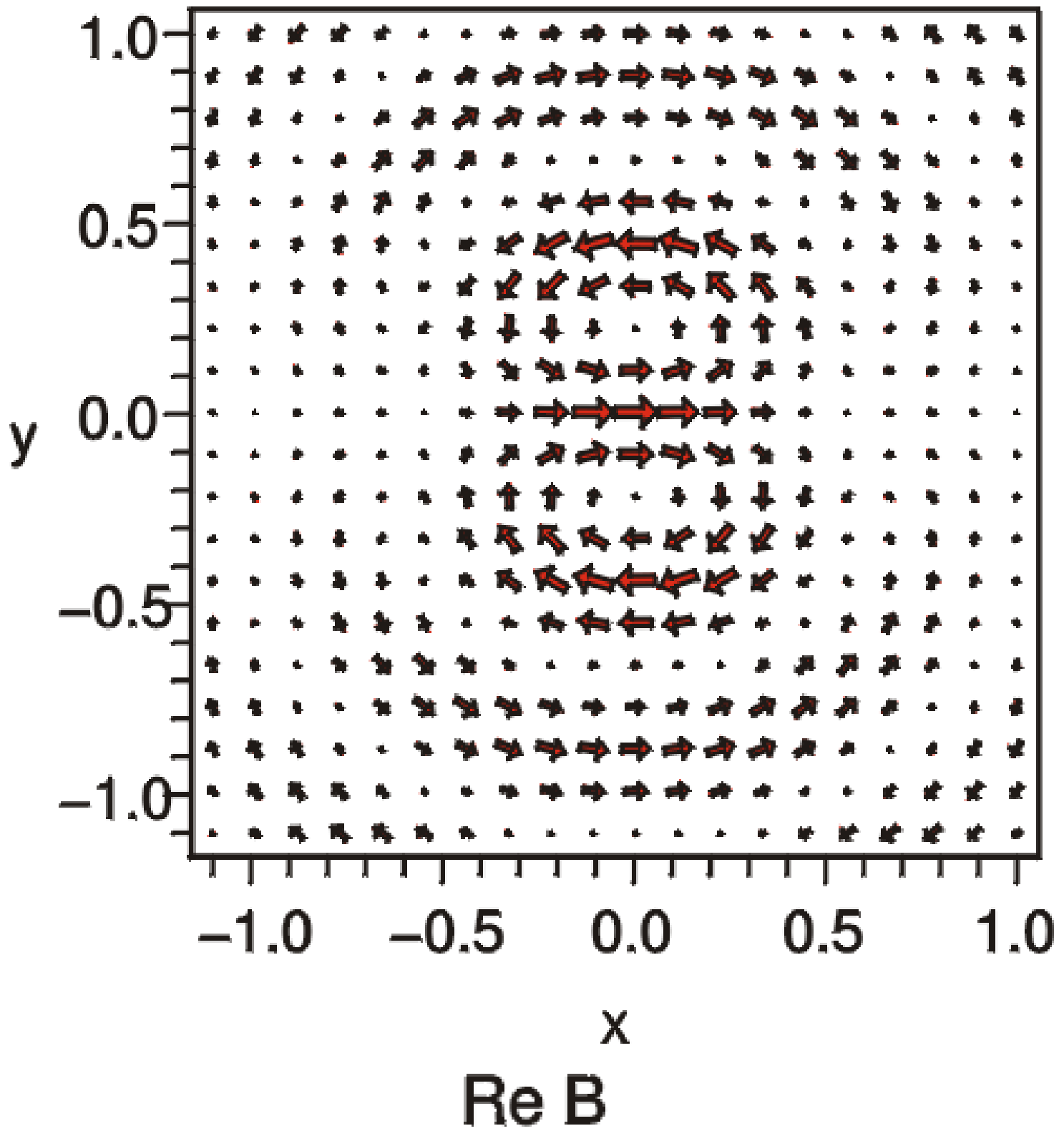

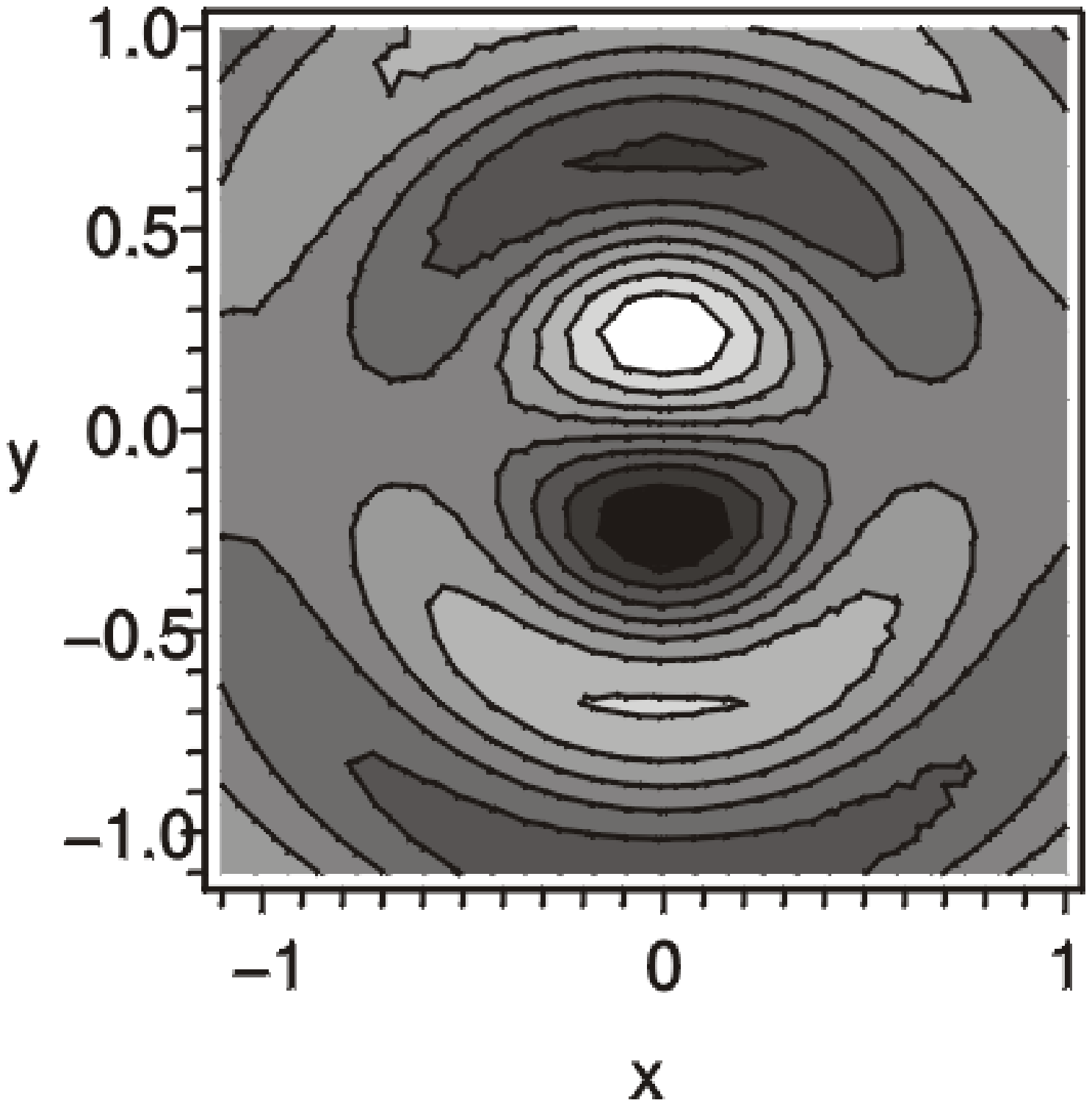

Fig.6.Spatial distribution of real part magnetic field in Laguerre-Gauss beams( $kw_0$=6,p=6,m=2 circular polarization) a) x-y components, b) z- component

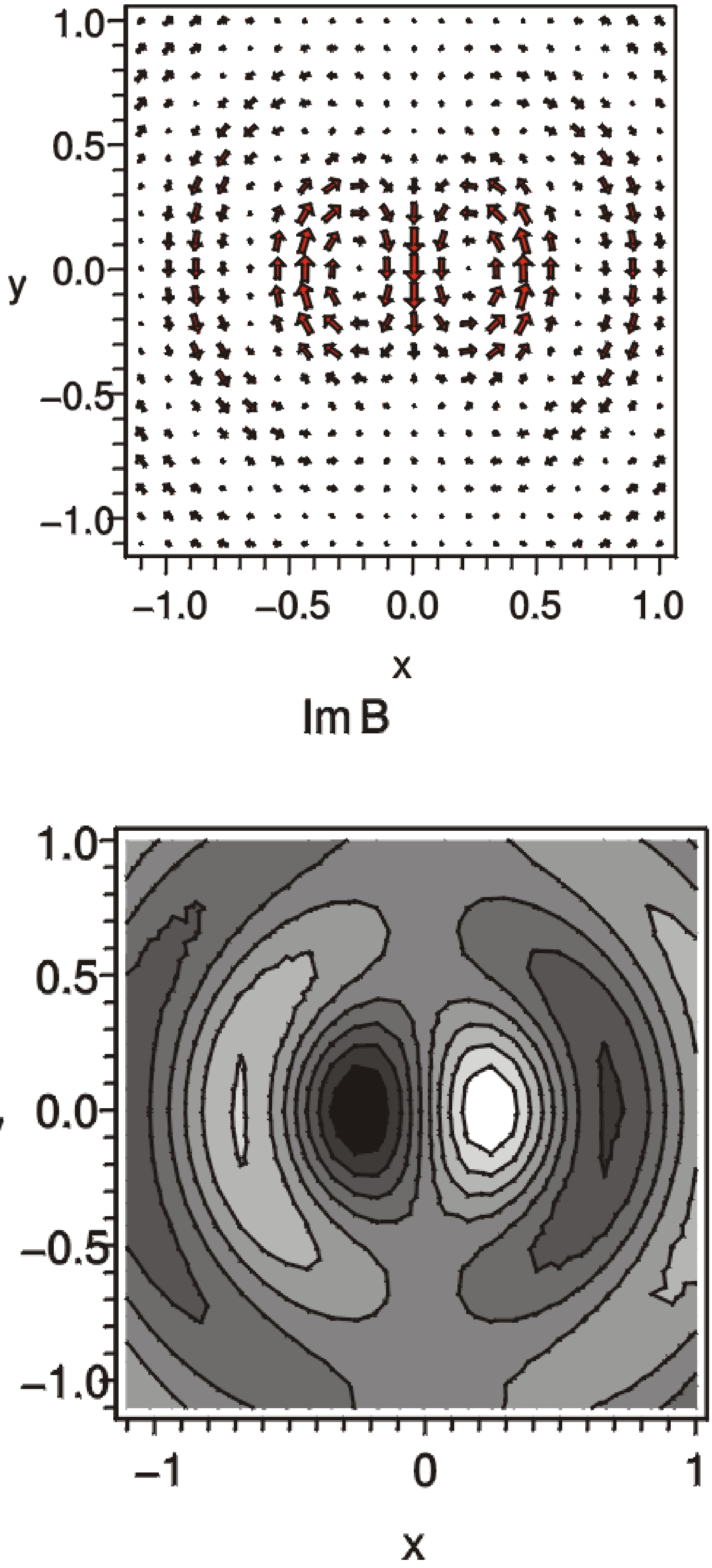

Fig.7.Spatial distribution of imaginary part magnetic field in Laguerre-Gauss beams( $kw_0$=6,p=6,m=2 circular polarization) a) x-y components, b) z- component

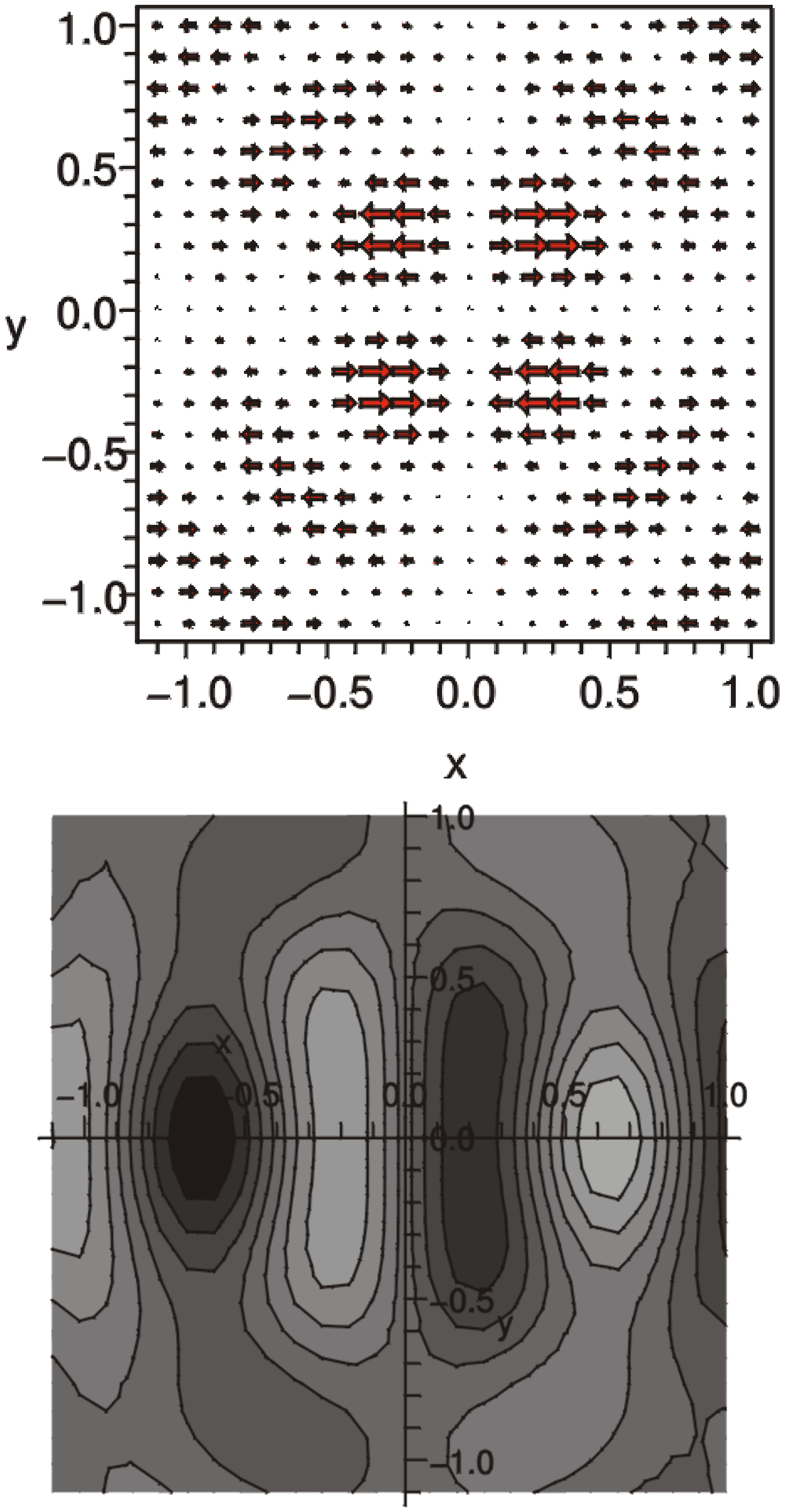

Fig.8.Spatial distribution of real part of electric field in Laguerre-Gauss beams ( $kw_0$=6,p=6,m=2,linear polarization ). a) x-y components, b) z- component

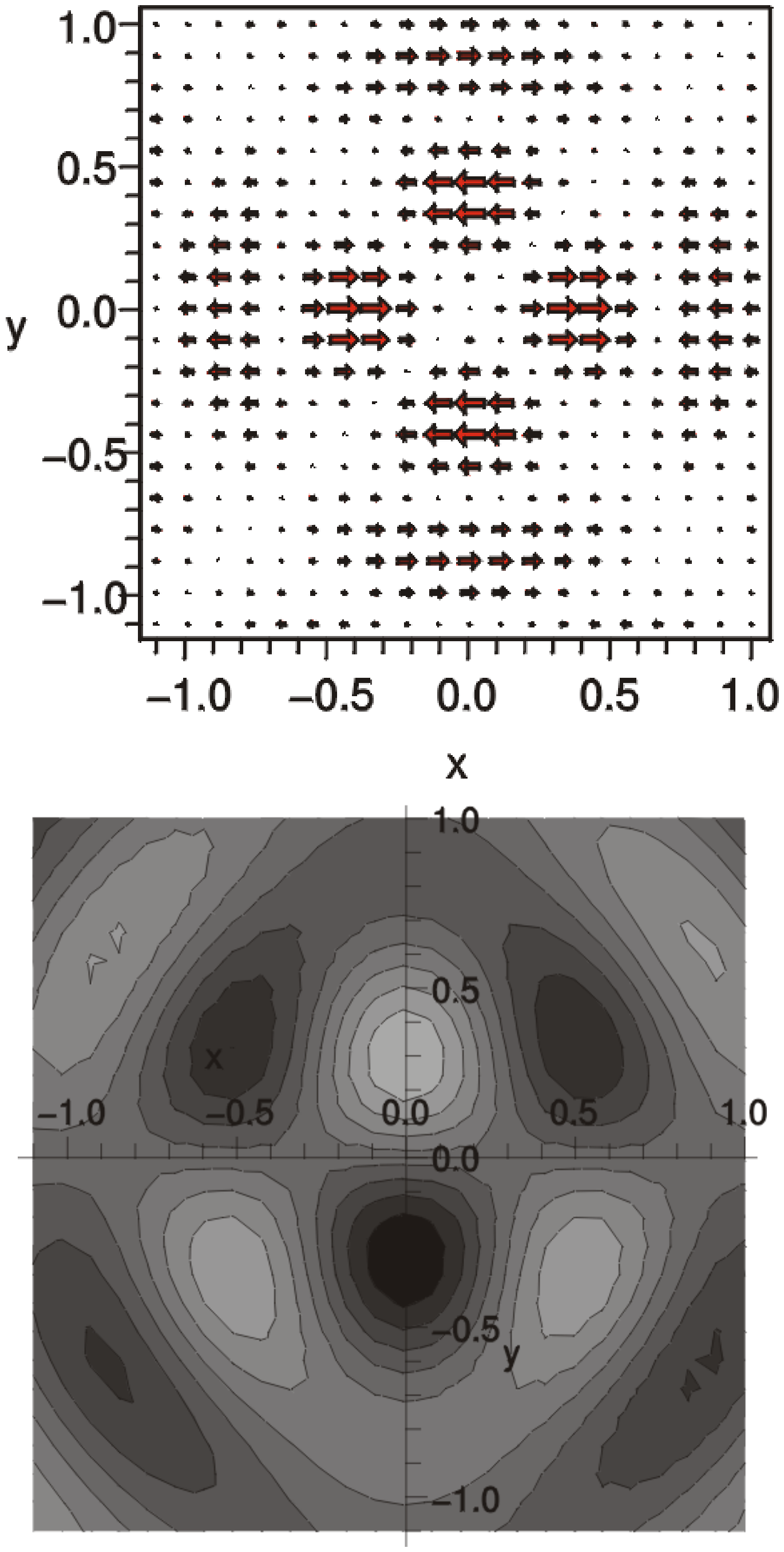

Fig.9.Spatial distribution of imaginary part of electric field in Laguerre-Gauss beams ( $kw_0$=6,p=6,m=2 linear polarization) a) x-y components, b) z- component

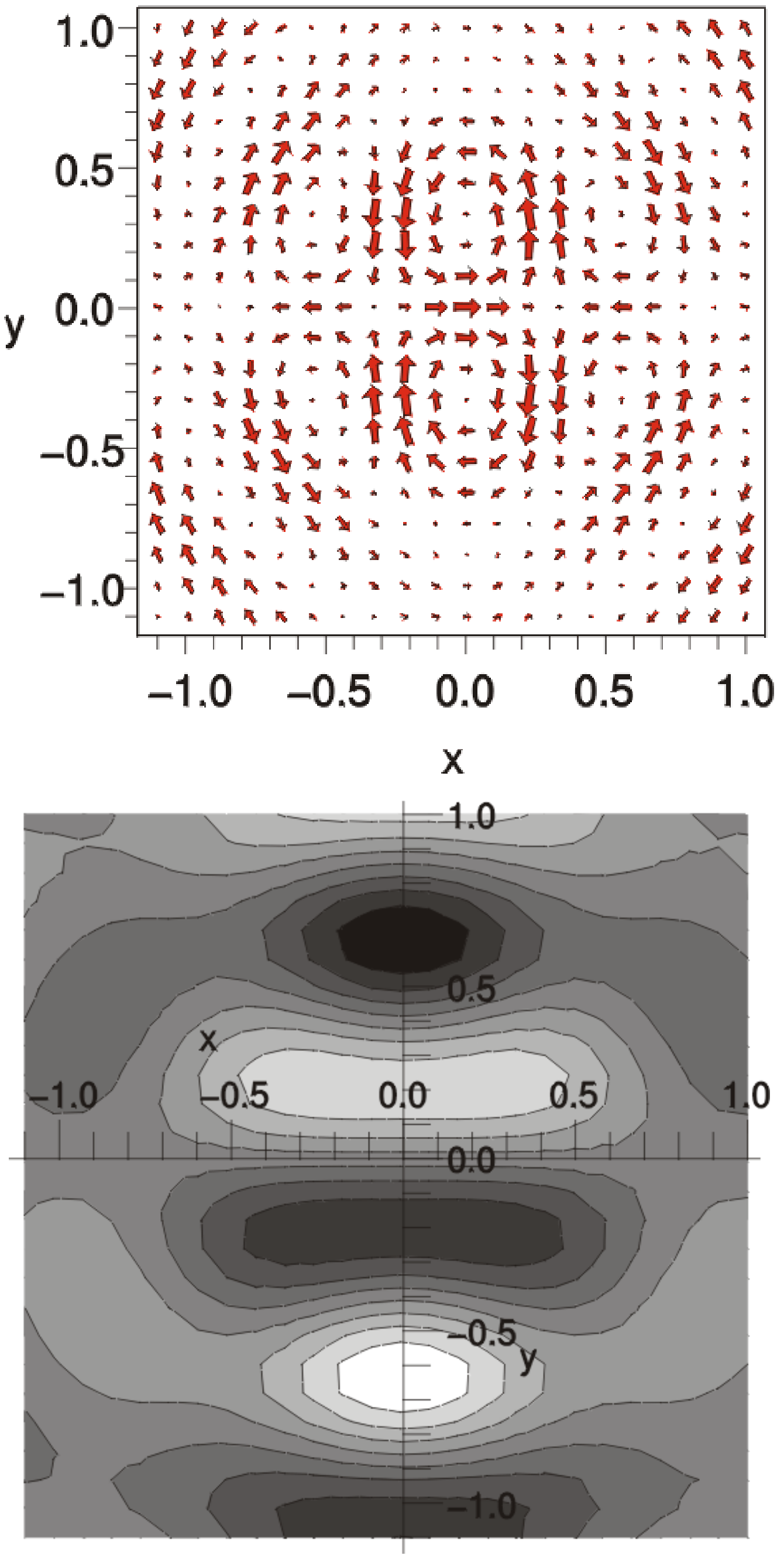

Fig.10.Spatial distribution of real part magnetic field in Laguerre-Gauss beams( $kw_0$=6,p=6,m=2 linear polarization) a) x-y components, b) z- component

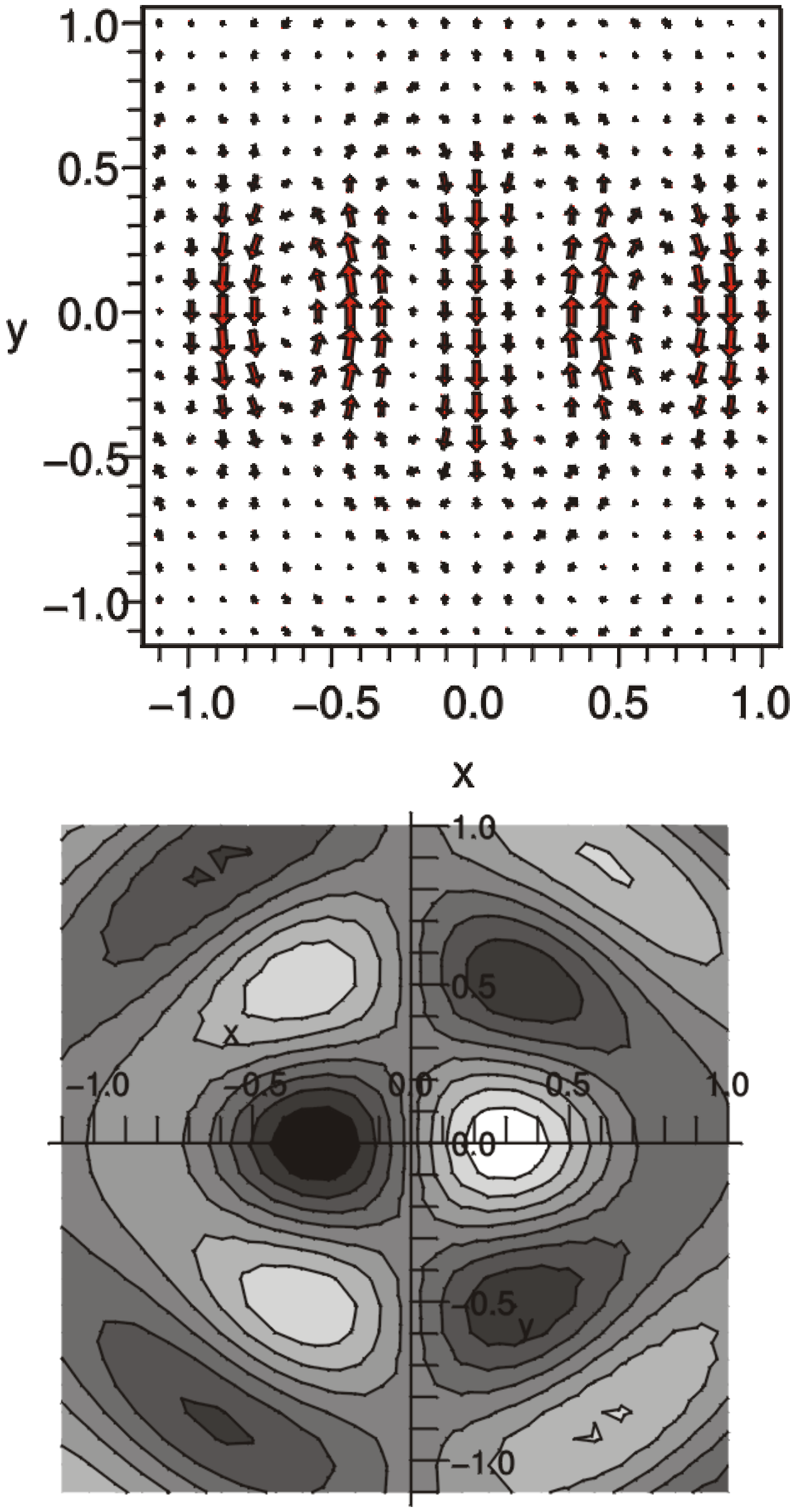

Fig.11. Spatial distribution of imaginary part magnetic field in Laguerre-Gauss beams( $kw_0$=6,p=6,m=2 linear polarization) a) x-y components, b) z- componentt

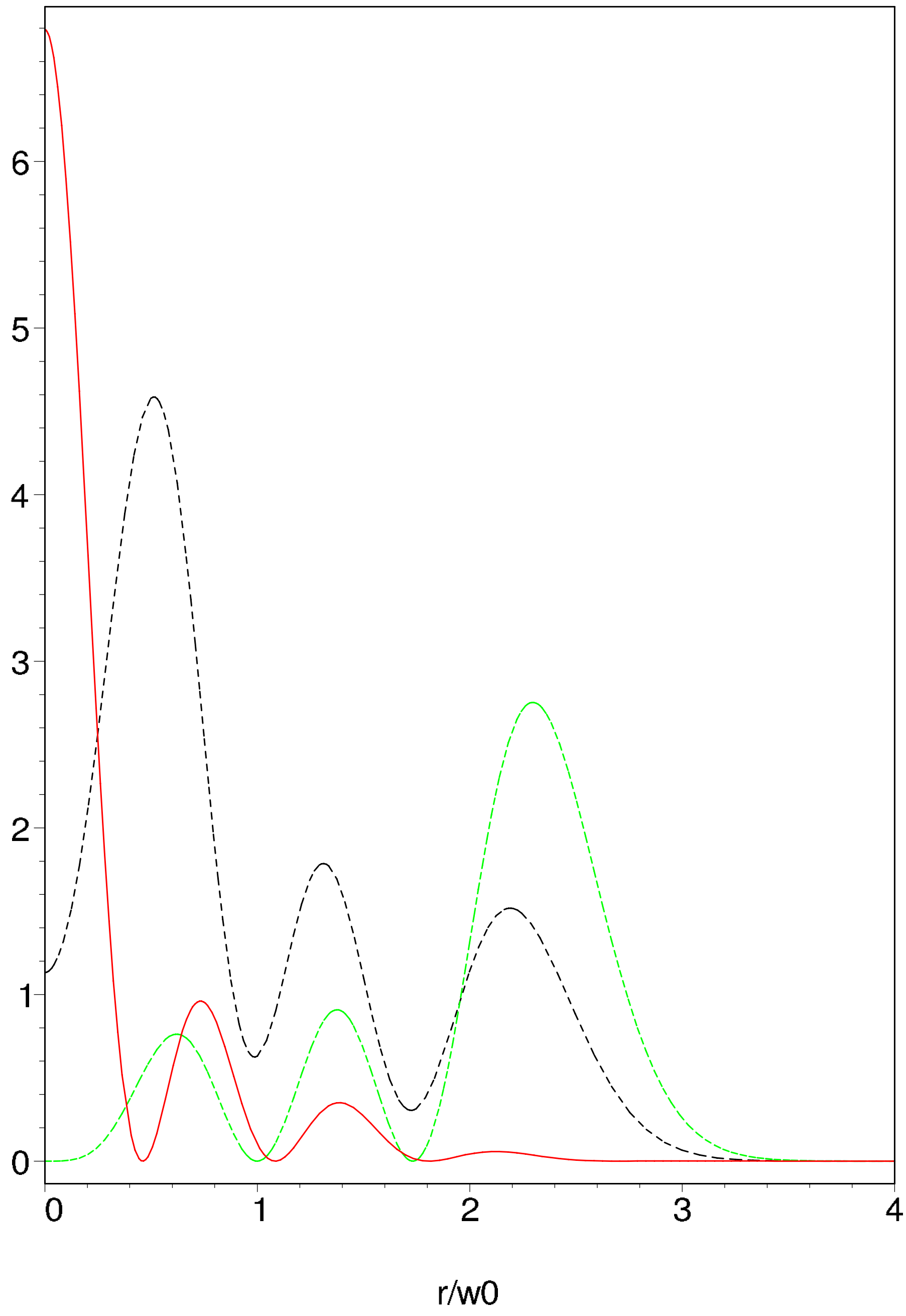

Fig.12. Normalized $\left(T_{E2}^{mM}/\left(E_0 Q^M\right)\right)^2$ radial distribution of excitation rates of E2(M=±1) transitions in detector placed in the waist of LG beams with p=2,m=2, $kw_0$=6. (M=+1 and M=-1 rates are shown by red and green lines). Dashed black line shows normalized distribution of full energy density.

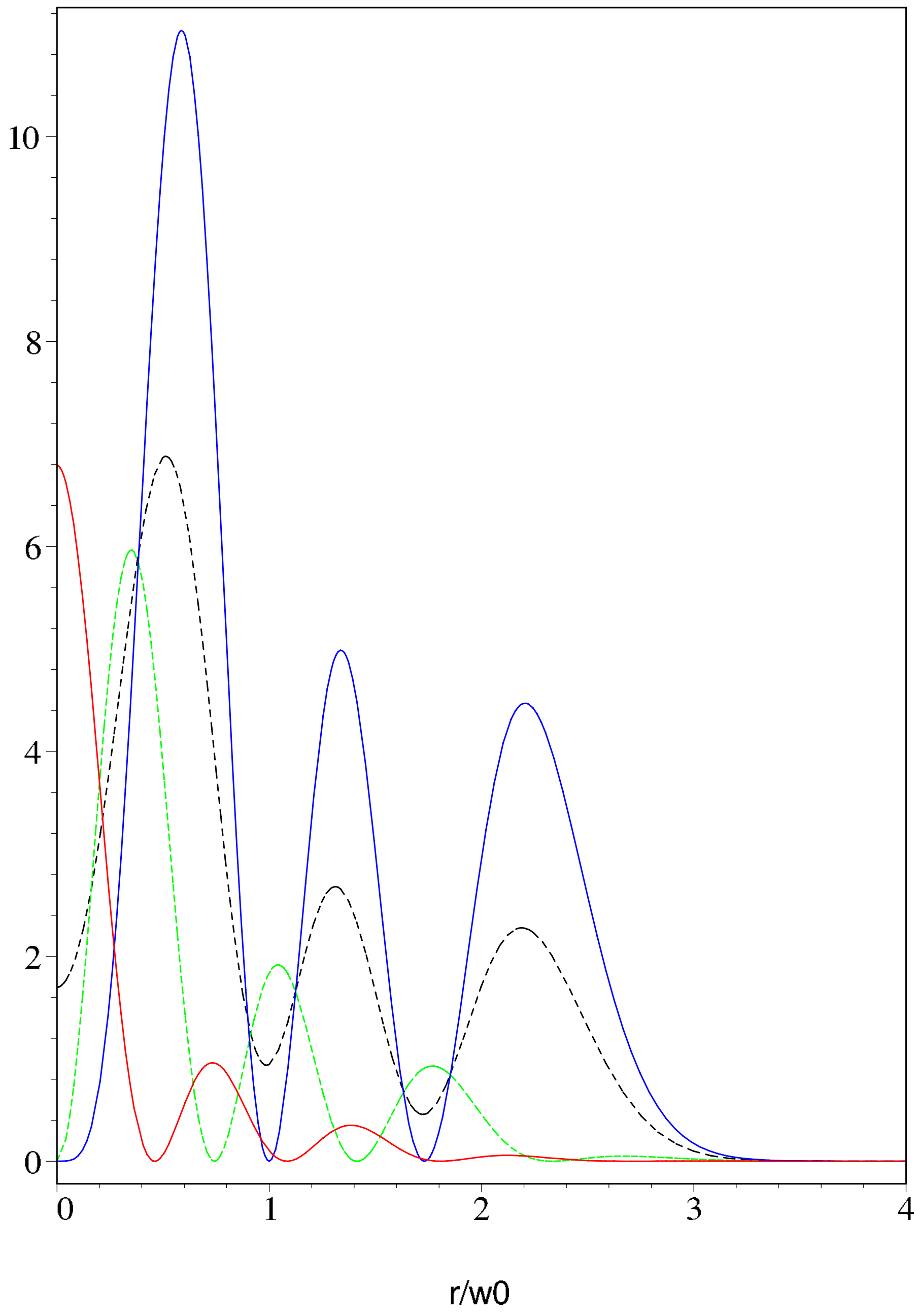

Fig.13.Normalized $\left(T_{M1}^{mM}/\left(E_0 m^M\right)\right)^2$ radial distribution of excitation rates of M1(M=±1) transitions in detector placed in the waist of LG beams with p=2,m=2, $kw_0$=6. (M=+1 , M=0, M=-1 are shown by red ,green and blue lines respectively ). Black line shows normalized distribution of full energy density

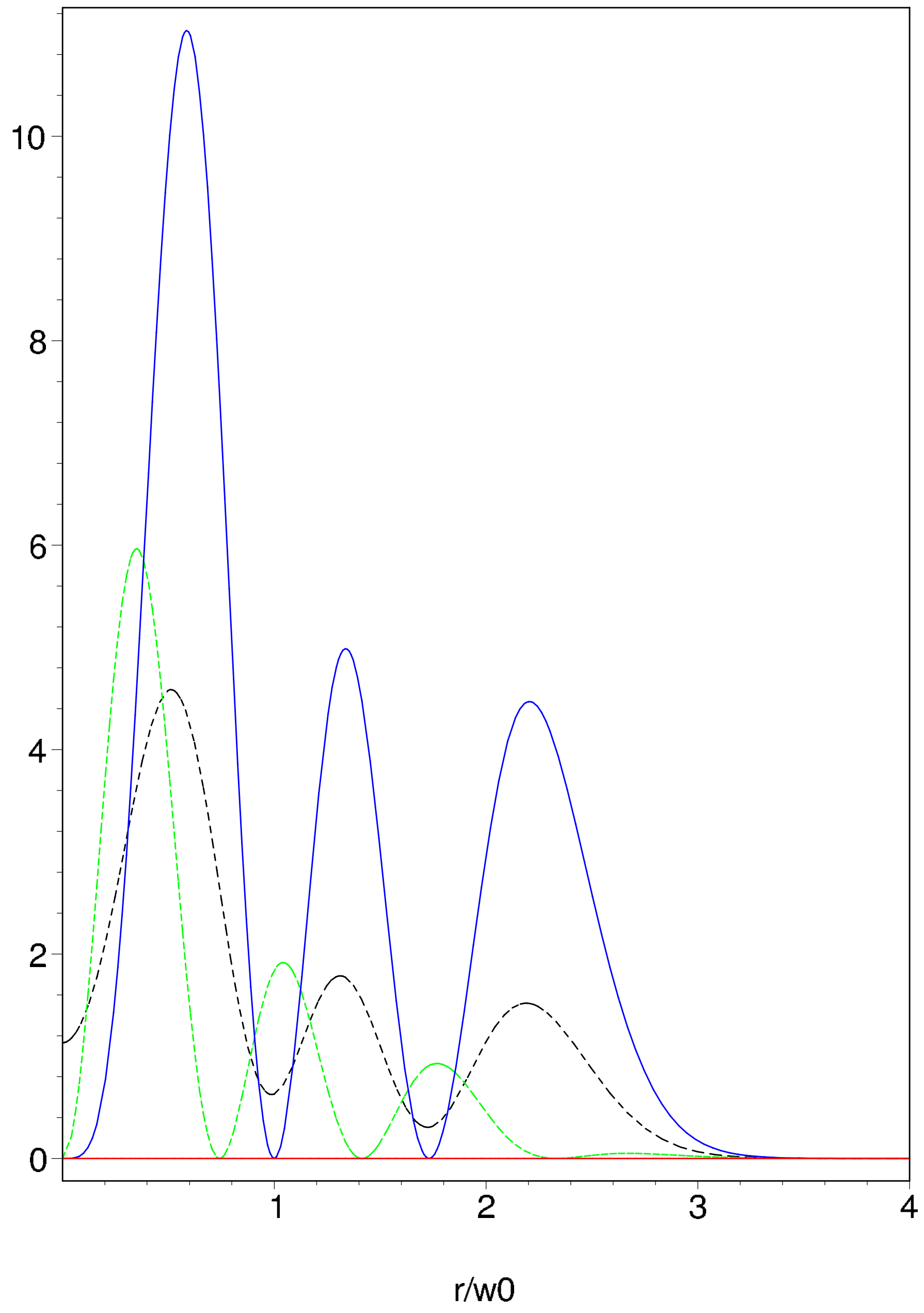

Fig.14. Normalized ($\left(T_{E1}^{mM}/\left(E_0 d^M\right)\right)^2$)radial distribution of excitation rates of E1(M=±1,0) transitions in detector placed in the waist of LG beams with p=2,m=2, $kw_0$=6. (M=+1 , M=0, M=-1 are shown by red ,green and blue lines respectively ). Black line shows normalized distribution of full energy density